%% file: main.tex
\documentclass[fleqn,usenatbib]{mnras}

\usepackage[T1]{fontenc}

\usepackage{graphicx}	
\usepackage{amsmath}	
\usepackage{amssymb}	
\usepackage{supertabular,booktabs}

\usepackage{newtxtext,newtxmath}

\DeclareRobustCommand{\VAN}[3]{#2}
\let\VANthebibliography\thebibliography
\def\thebibliography{\DeclareRobustCommand{\VAN}[3]{##3}\VANthebibliography}

\DeclareRobustCommand{\VANDER}[3]{#2}
\let\VANDERthebibliography\thebibliography
\def\thebibliography{\DeclareRobustCommand{\VANDER}[3]{##3}\VANDERthebibliography}

\def\paper1{FR22}

\title[Eccentric planet-disc interactions]{Eccentric planet-disc interactions: orbital migration and eccentricity evolution}

\author[Fairbairn \& Rafikov]{
Callum W. Fairbairn$^{1}$\thanks{E-mail: cfairbairn@ias.edu}
and Roman R. Rafikov$^{1,2}$\thanks{E-mail: rrr@damtp.cam.ac.uk}
\\
$^{1}$School of Natural Sciences, Institute for Advanced Study\\
1 Einstein Drive, Princeton 08540, NJ\\
$^{2}$Department of Applied Mathematics and Theoretical Physics, University of Cambridge, Centre for Mathematical Sciences,\\
Wilberforce Road, Cambridge CB3 0WA, UK
}

\date{Accepted XXX. Received YYY; in original form ZZZ}

\pubyear{2024}

\begin{document}
\label{firstpage}
\pagerange{\pageref{firstpage}--\pageref{lastpage}}
\maketitle

\input{Sections/0_abstract.tex}




\input{Sections/1_introduction.tex}

\input{Sections/2_linear_framework.tex}

\input{Sections/3_circular_benchmark.tex}
\input{Sections/4_eccentric_torques.tex}
\input{Sections/5_discussion.tex}

\input{Sections/6_conclusions.tex}

\input{Sections/acknowledgements}

\section*{Data Availability}

The data underlying this article are available in the online supplementary material, as described in Appendix \ref{app:parameter_space}.



\bibliographystyle{mnras}
\bibliography{references.bib} 



\appendix
\input{Sections/app_convergence}
\input{Sections/app_spurious_modes}

\input{Sections/app_parameter_space}



\bsp	
\label{lastpage}
\end{document}

%% file: Sections/0_abstract.tex
\begin{abstract}
Gravitational coupling between a protoplanetary disc and an embedded eccentric planet is an important, long-standing problem, which has been not yet been conclusively explored. Here we study the torque and associated orbital evolution of an eccentric planet in a two-dimensional disc via the semi-analytical, fully global linear approach. Our methodology has the advantage that the spatial structure of the density waves launched by the planet is solved for fully. This allows us to account for the possibility of torque excitation over an extended radial interval for each Fourier harmonic of the perturbation, as opposed to earlier approximate treatments localized around Lindblad and corotation resonances. We systematically explore the torque behaviour across the space of disc properties (assuming power law profiles for the disc surface density and temperature), including the aspect ratio. Crucially, we examine the torque variation as the orbital eccentricity becomes supersonic relative to the gas motion (when planetary eccentricity is of order the disc aspect ratio), finding that the torque robustly reverses its sign near this transition. We also find that for shallow surface density gradients planetary migration may become outwards beyond this transition, although the rapid eccentricity damping (which is typically $\sim 10^2$ times faster than the orbital migration rate) would quickly restore inwards migration as the planet circularizes. Our self-consistently computed torques are in qualitative agreement with past numerical studies of eccentric planet-disc coupling. We provide our torque data for different disc parameters to the community for future testing and implementation in population synthesis studies.
\end{abstract}

%% file: Sections/1_introduction.tex
\section{Introduction}
\label{sec:background}

Gravitational coupling between a gaseous disc and an embedded massive perturber --- a planet or a satellite --- is known to be one of the natural drivers of the protoplanetary disc evolution and a key factor determining the architectures of exoplanetary systems \citep{Paardekooper20232023}. Since the early pioneering studies of \citet{GoldreichTremaine_1979,GoldreichTremaine_1980} and \citet{Lin1979} it was known that the exchange of angular momentum between a planet and its nascent disc can lead to planet migration and may truncate the disc at certain locations leading to gap opening. The latter effect may ultimately be responsible for the origin of various substructures that have been observed in protoplanetary discs through e.g. thermal dust emission \citep{AndrewsEtAl_2018,Andrews2020}.

It is also well known observationally that orbits of many exoplanets have appreciable eccentricities \citep{Santos2007,Xie2016}. While the mechanisms and timing of the eccentricity production are still debated \citep{Davies2014}, it is certainly plausible that planets could have a significant eccentricity while still embedded in protoplanetary discs, e.g. due to planet-planet scattering\footnote{More massive planets capable of opening clean gaps may develop non-zero eccentricity also due to gravitational planet-disc coupling \citep{Goldreich2003,Ragusa2018}.}. This naturally raises a possibility of gravitational coupling between the disc and an embedded eccentric planet, a topic on which a number of results have been obtained since the early days in the limit of small planetary eccentricity \citep{GoldreichTremaine_1980,Ward_1986,Artymowicz_1993}. However, the approximations often made in these studies and the differences in their predictions \citep{PapaloizouLarwood_2000,CresswellNelson_2006,IdaEtAl_2020} warrant a re-evaluation of this problem using a first-principles calculation. This is what we do in this work, focusing on the limit of low mass planets to make our calculations rigorous.

Planet-disc coupling is a two-stage process mediated by the density waves driven in the disc by the planet. In the first stage the non-axisymmetric planetary potential excites a density wave, injecting angular momentum into the wave via the {\it excitation} torque acting on the perturbed surface density in the disc. From the global angular momentum conservation, an equal and opposite torque must act on the planet, driving the evolution of its orbital elements --- semi-major axis and eccentricity (in the planar case). In the second stage, the wave damps as it propagates through the disc, either linearly \citep{Takeuchi1996,MirandaRafikov_2020} or non-linearly \citep{GoodmanRafikov_2001,Rafikov2002a}, ultimately depositing its angular momentum into the disc fluid. It is this latter process that causes the disc to evolve, leading to gap opening and the emergence of various observable substructures. 

In the present work (in line with many existing studies) we will be interested only in the former, wave excitation, process, which is important to understand even when it is considered in isolation. First, in most cases the excitation torque fully determines the orbital evolution of planets, which is the main motivation for this study. Second, the excitation torque should be quite insensitive to the details of the subsequent wave damping unless the wave dissipation is very fast (a possibility which we do not consider in this study). For brevity, in the following we will refer to the excitation torque simply as `torque'. There are multiple ways\footnote{Some studies, e.g. \citet{IdaEtAl_2020}, have synthesized orbital evolution prescriptions using multiple approaches listed below.} in which planetary torque and the evolution of planetary orbital elements can be explored. 

First, they can be extracted directly from numerical simulations of eccentric planet-disc interaction, either in 2D or 3D geometry \citep{CresswellNelson_2006,CresswellNelson_2008,Cresswell2007} by following the obital evolution of simulated planets gravitationally interacting with the disc. 

Second, there have been attempts to compute the planetary torque by considering the force acting on the planet as a variant of the dynamical friction against the gaseous background of the underlying disc \citep{MutoEtAl_2011,Desj2022,BuehlerEtAl_2024,ONeill2024}. Unfortunately, to make the calculation tractable these studies have to assume an idealized unperturbed background state which differs from the sheared Keplerian flow in a real disc. 

Third, many studies of the planetary orbital evolution employ a Fourier-space based approach rooted in the analytical methodology of the torque calculation proposed in the seminal works of \citet{GoldreichTremaine_1979,GoldreichTremaine_1980} and \citet{Lin1979}. The idea lies in linearization and (azimuthal) Fourier decomposition of the fluid equations governing planet-disc coupling, with the resultant equations possessing a number of singularities at the radial locations corresponding to the Lindblad and corotation resonances (see Section \ref{subsection:resonances}). A subsequent simplification of these equations in the vicinity of these resonances in the WKB limit shows that (1) the torque is excited only at these resonances and (2) its amplitude can be related in a very simple way to the planetary potential (and its radial derivative) evaluated at the resonance, {\it without actually solving} the original equations for fluid perturbations. Subsequent studies introduced various modifications to this prescription, which incorporated the leading order background gradients in the disc \citep{Ward_1986,Ward_1988}, abandoned the WKB assumption and reintroduced the effect of azimuthal pressure gradients for large azimuthal wavenumbers \citep{Artymowicz_1993, Artymowicz1993b, Artymowicz_1994, Ward_1997}. The latter refinements modified the torque formula to sample the perturbing potential over a finite radial interval around the resonance (rather than at its exact location) and allowed it to account for a \textit{torque cut-off} \citep[first described in][]{GoldreichTremaine_1980} which quenches the torque contributions from the highest order modes. Unfortunately, results obtained with this approach are only approximate even in the linear limit as a result of local, near-resonance expansion of the governing equations. Furthermore, they do not easily yield the radial distribution of the planetary torque across the disc \citep[only its integrated value, cf.] []{RafikovPetrovich_2012}, and typically describe planetary eccentricity evolution only in the limit $e\to 0$, except for \citet{PapaloizouLarwood_2000}.

Fourth, an alternative and perhaps somewhat under-appreciated approach is to forgo any simplifications of the linearized and Fourier-decomposed equations altogether and to pursue their direct numerical solution in 2D, as first performed by \cite{KorycanskyPollack_1993} for planets on circular orbits. This calculation has also been extended to 3D in \citet{TanakaEtAl_2002}, \citet{TanakaWard_2004}, \citet{TanakaOkada_2024}. In the linear limit this semi-analytical approach should give an {\it exact} solution for the planetary torque \citep{RafikovPetrovich_2012,Petrovich2012} and the Fourier mode structure, allowing one to reconstruct the full spatial distribution of the planet-driven perturbation \citep{MirandaRafikov_2019a,MirandaRafikov_2020}.

In this paper we follow \citet{KorycanskyPollack_1993} and employ this latter, semi-analytical approach, making no assumptions or simplifications apart from the linearity of planet-disc coupling, to explore the orbital evolution of planets in a 2D disc over a range of planetary eccentricity, not necessarily assuming $e$ to be very small. In particular, this allows us to probe a problematic regime of transonic planetary motion for which there is some controversy in the existing results \citep{PapaloizouLarwood_2000,CresswellNelson_2006,IdaEtAl_2020}; namely when $e$ is of the order of the disc aspect ratio. We explore the transonic transition over a whole grid of disc parameters --- aspect ratio, surface density and temperature slopes --- and provide accurate prescriptions for the planetary migration and eccentricity evolution timescales as a function of these parameters and eccentricity. 

Our calculations utilize the semi-analytic framework for studying the eccentric planet-disc coupling that was developed in \citet[][hereafter \paper1]{FairbairnRafikov_2022} to explore the spatial structure of the perturbation induced in a 2D disc by an eccentric planet. This framework was shown to provide a very good match for the time-resolved morphology of the wake induced in the disc by an eccentric planet that was previously investigated via direct hydrodynamic simulations by \citet{ZhuZhang_2022}. Whilst in \paper1 we focused on synthesising the wake morphology, we now exploit this framework to investigate the planet-disc torque acting on an eccentric planet and to provide prescriptions for the evolution of the planetary orbit. 

This paper is organized as follows. In Section \ref{sec:linear_method} we will summarise the prerequisite linear theory and numerical pipeline developed in \paper1. In Section \ref{sec:circular_benchmark} we benchmark this procedure on circular orbits before turning our attention to eccentric orbits in Section \ref{sec:eccentric_torques}. We will discuss these results in the context of previous studies in Section \ref{sec:discussion} before concluding in Section \ref{sec:conclusion}. Furthermore, Appendix \ref{app:parameter_space} describes how to access and process our data which can be used to easily obtain planetary migration and eccentricity evolution timescales over the range of parameters explored in this work.

%% file: Sections/2_linear_framework.tex
\section{Linear framework}
\label{sec:linear_method}

In this section we will summarise the linear procedure being used to study eccentric planet-disc interactions. We will begin by reviewing the orbital setup of the embedded planet and also the physical parameters governing the disc profile. We will then outline the linearised equations being solved and introduce the numerical mode finding method. For more details on the setup and an in-depth description of the solution techniques, we refer the reader to Sections 2 and 3 of \paper1 and the references therein.

\subsection{Planet-disc setup}
\label{subsection:planet_disc_setup}

We adopt a single planet with mass $M_\textrm{p}$ orbiting a host star of mass $M_{*}$. The anticipated migration timescale and evolution of orbital elements resulting from the planet-disc interaction are much longer than the orbital timescale, so we assume the planet is on a fixed, eccentric orbit \citep[as per][]{GoldreichTremaine_1980}. This orbit is parameterised by two elements, relative to a coordinate system centred on the star, planetary eccentricity $e$ and the semi-major axis $a$. The Kepler problem then dictates that the orbital mean motion of the planet is $\Omega_\textrm{p} = \sqrt{G M_{*}/a^3}$. The argument of periapse is arbitrarily chosen so it conveniently aligns with the coordinate frame $x$-axis. In this polar reference frame, the planetary radial coordinate is denoted $r_\textrm{p}$, whilst the true anomaly is given by $\phi_\textrm{p}$ (see Fig. 1 in \paper1). Meanwhile, an arbitrary position in the disc is labelled by the radial and azimuthal coordinates $r$ and $\phi$ respectively. 

The gravitational influence of the planet leads to a disturbing potential acting on the disc. In this paper we will consider only the \textit{direct} potential contribution which is given by
\begin{equation}
    \label{eqn:direct_potential}
    \Phi(\mathbf{r}) = -\frac{G M_\textrm{p}}{|\mathbf{r}-\mathbf{r}_\textrm{p}|} ,
\end{equation}
where $\mathbf{r}$ and $\mathbf{r}_\textrm{p}$ denote the arbitrary position vector relative to the star and the position vector of the planet respectively. Note that we do not account for the \textit{indirect} potential in this work. This contributes an additional term which is proportional to $\cos{\phi}$, owing to the non-inertial reference frame centred on the star rather than the planet-star barycenter. Previous authors often disregard this term \citep[e.g.][]{KorycanskyPollack_1993,PapaloizouLarwood_2000,TanakaEtAl_2002}, so for compatibility when comparing our results, this has also been neglected here. Indeed, \cite{MirandaRafikov_2019a} and \paper1 found that inclusion of the indirect term does not significantly impact the wake morphology.

The second key component of the problem is the gaseous disc with which the embedded protoplanet interacts. We will consider a 2D, coplanar, inviscid disc model which is perhaps the simplest scenario but still contains a richness of physics. The unperturbed, axisymmetric disc is described by smooth power-law profiles in the surface density
\begin{equation}
    \label{eqn:surface_density}
    \Sigma(r) = \Sigma_{\textrm{p}}\left(\frac{r}{a}\right)^{-p} ,
\end{equation}
and the locally isothermal sound speed 
\begin{equation}
    \label{eqn:sound_speed}
    c_\textrm{s} (r) = c_{\textrm{s},\textrm{p}}\left(\frac{r}{a}\right)^{-q/2}.
\end{equation}
We adopt a simple locally isothermal equation of state (EoS) which then sets the vertically integrated pressure $P$ according to 
\begin{equation}
    \label{eqn:eos}
    P = c_{\textrm{s}}^2 \Sigma ,
\end{equation}
such that there is a radial temperature profile
\begin{equation}
    \label{eqn:temperature}
    T_\textrm{disc}(r) \propto \frac{P}{\Sigma} = c_\textrm{s}^2 \propto r^{-q},
\end{equation}
controlled by the exponent $q$. This choice for the EoS affects the form of the underlying perturbation equations and the resulting wave propagation behaviour as shown by \cite{MirandaRafikov_2020}. According to hydrostatic equilibrium, the aspect ratio $h(r)$ of the disc then follows
\begin{equation}
    \label{eqn:aspect_ratio}
    h(r) \equiv \frac{H(r)}{r} = h_\textrm{p} \left(\frac{r}{a}\right)^{(1-q)/2},
\end{equation}
where $H(r) = c_s(r)/\Omega_\textrm{K}(r) $ is the pressure scale height and $\Omega_\textrm{K}(r) = \sqrt{GM_{*}/r^3}$ denotes the local Keplerian velocity. Note that all of the quantities subscripted by `$\textrm{p}$' are simply characteristic values evaluated at the planetary semi-major axis $a$. 

Associated with the surface density and temperature structure is an additional pressure support in the radial force balance for the disc. This leads to a slightly sub-Keplerian fluid orbital frequency $\Omega$ due to order $h_{\textrm{p}}^2$ corrections:
\begin{equation}
    \label{eqn:fluid_frequency}
    \Omega^2(r) = \Omega_\textrm{K}^2+\frac{1}{r \Sigma}\frac{dP}{dr}=\Omega_\textrm{K}^2\left[1-h_\textrm{p}^2(p+q)\left(\frac{r}{a}\right)^{1-q}\right].
\end{equation}
Similarly the fluid epicylic frequency $\kappa$ is slightly detuned from the Keplerian value $\Omega_\textrm{K}$ and is given by
\begin{equation}
    \kappa^2 = \frac{2\Omega}{r}\frac{d}{dr}\left(r^2\Omega\right) = \Omega_\textrm{K}^2\left[1-h_\textrm{p}^2(p+q)(2-q)\left(\frac{r}{a}\right)^{(1-q)}\right].
\end{equation}

\subsection{Linear Equations}
\label{subsection:linear_equations}

Linear theory has paved the way for understanding planet-disc interaction since the seminal work of \cite{GoldreichTremaine_1979, GoldreichTremaine_1980}. To be more specific, linearity requires that $M_\textrm{p} \lesssim M_\textrm{th}$ where the thermal mass is defined as $M_\textrm{th} = h_\textrm{p}^3 M_*$. This condition is equivalent to requiring that the Hill radius of the planet is less than the local pressure scale height of the disc.

The mass continuity and Euler equations can then be linearised by decomposing the fluid variables into the steady background and perturbed components in accordance with $X\rightarrow X+\delta X$ (dropping subscript `$0$' from the unperturbed variables). Notably the direct planetary potential acts as a perturbing source term which admits the Fourier expansion
\begin{equation}
    \label{eqn:potential_expansion}
    \Phi = \sum_{m = 1}^{\infty} \sum_{l = -\infty}^{\infty} \Phi_{ml}(r)\cos\left(m\phi-l \Omega_\textrm{p} t\right) ,
\end{equation}
where only the cosine term appears since we are free to choose a time origin such that the planet is at $\phi = 0$ when $t=0$. The presence of two modal numbers $m$ and $l$ indicates that there are two fundamental periodicities in the problem. Whereas a circular planet appears stationary in a frame corotating with the mean motion, and only requires the azimuthal quantum number $m$, an eccentric planet will execute epicylic oscillations in the frame aligned with the planetary guiding centre and hence require a temporal expansion in $l$ also. This form shows that each $(m,l)$ modal combination contributes an $m$-armed potential forcing with a pattern speed
\begin{equation}
    \label{eqn:pattern_speed}
    \omega_{ml} = \frac{l}{m}\Omega_\textrm{p}.
\end{equation}
This forcing will drive a corresponding Fourier modal response in the fluid perturbations which are also expanded according to 
\begin{equation}
    \label{eqn:modal_repsonse}
    \delta X = \mathrm{Re}\left[\delta X(r)e^{\mathrm{i}(m\phi-l\Omega_\textrm{p}t)} \right],
\end{equation}
where for brevity we have dropped subscripts $(m,l)$ from $\delta X_{ml}$. Inserting this ansatz into the linear equations reduces the partial differential equations to a set of coupled ordinary differential equations governing the radial mode structure:
\begin{align}
\label{eq:density_pert_eqn}
&    -i \Tilde{\omega}\delta \Sigma + \frac{1}{r}\frac{d}{d r}(r\Sigma \delta u_r)+\frac{i m \Sigma}{r}\delta u_\phi = 0 , 
    \\
\label{eq:ur_eqn}
&    -i \Tilde{\omega}\delta u_r - 2\Omega \delta u_{\phi} = -\frac{1}{\Sigma}\frac{d}{d r} \delta P +\frac{1}{\Sigma^2}\frac{dP}{dr}\delta \Sigma-\frac{d}{d r}\Phi_{ml},
    \\
\label{eq:uphi_eqn}
& -i \Tilde{\omega}\delta u_\phi + \frac{\kappa^2}{2\Omega}\delta u_r = -\frac{i m}{r}\left( \frac{\delta P}{\Sigma}+\Phi_{ml}\right).
\end{align}
Here $\tilde{\omega} = m(\omega_{ml}-\Omega)$ is the Doppler-shifted forcing frequency experienced by a fluid element in the disc as the perturbing potential mode sweeps by. These equations can be combined along with the EoS \eqref{eqn:eos} to yield the master equation
\begin{align}
\label{eq:master_eqn}
     \frac{d^2}{dr^2}\delta h & +\left\{ \frac{d}{dr}\ln\left(\frac{r\Sigma}{D}\right)-\frac{1}{L_T} \right\} \frac{d}{dr}\delta h \nonumber \\
    & - \left\{ \frac{2m\Omega}{r\Tilde{\omega}}\left[\frac{1}{L_T}+\frac{d}{dr}\ln\left(\frac{\Sigma \Omega}{D}\right)\right] \right. \nonumber \\
    & \left.+\frac{1}{L_T}\frac{d}{dr}\ln\left(\frac{r\Sigma}{L_T D}\right)+\frac{m^2}{r^2}+\frac{D}{c_{\textrm{s}^2}}\right\} \delta h = \nonumber \\
    & = -\frac{d^2 \Phi_{ml}}{dr^2} - \left[ \frac{d}{dr}\ln\left(\frac{r\Sigma}{D}\right)\right]\frac{d\Phi_{ml}}{dr} \nonumber \\
    & + \left\{\frac{2m\Omega}{r\Tilde{\omega}}\left[\frac{d}{dr}\ln\left(\frac{\Sigma \Omega}{D}\right)\right]+\frac{m^2}{r^2}\right\}\Phi_{ml},
\end{align}
in terms of the pseudo-enthalpy variable $\delta h = \delta P/\Sigma$. Here $L_\textrm{T}$ denotes the characteristic locally isothermal length scale set by
\begin{equation}
    \label{eqn:thermal_scale}
    \frac{1}{L_\textrm{T}} = \frac{d \ln c_\textrm{s}^2}{dr} ,
\end{equation}
whilst $D$ measures the offset from a resonance given by
\begin{equation}
    \label{eqn:resonance_offset}
    D = \kappa^2-\tilde{\omega}^2.
\end{equation}
This formulation of the problem looks equivalent to equations (17)-(21) presented by \cite{MirandaRafikov_2020}, but now $\tilde{\omega}$ and $D$ crucially involve the different pattern speeds $\omega_{ml}$ corresponding to the eccentric planetary motion. When later computing the torque and angular momentum flux, we will use the original fluid variables which are given by  
\begin{align}
    \label{eq:dsigma_dh}
    \delta\Sigma & = \frac{\Sigma \delta h}{c_{\textrm{s}^2}} , \\
    \label{eq:dur_dh}
    \delta u_r &= \frac{i}{D} \left[ \left( \Tilde{\omega}\frac{d}{d r}-\frac{2m\Omega}{r}\right)(\delta h+\Phi_{ml})-\frac{\Tilde{\omega}}{L_T}\delta h\right] , \\
    \label{eq:duphi_dh}
    \delta u_\phi & = \frac{1}{D} \left[ \left( \frac{\kappa^2}{2\Omega}\frac{d}{d r}-\frac{m\tilde{\omega}}{r}\right)(\delta h+\Phi_{ml})-\frac{\kappa^2}{2\Omega L_T}\delta h \right].
\end{align}
%

\subsection{Review of resonances}
\label{subsection:resonances}

Various singularities are present in equation \eqref{eq:master_eqn} at the radii where $\omega_{ml}=\Omega$ (i.e. $\tilde\omega=0$) or $D=0$, which evidences the existence of resonant locations. Physically they correspond to a maximal coupling between the forcing due to the perturbing potential and the free oscillations supported by the disc -- thereby yielding sites where waves are launched and angular momentum is efficiently transferred between the planet and fluid \citep[e.g.][]{GoldreichTremaine_1979}. These resonances come in a variety of flavours. 

\subsubsection{Lindblad resonances}

Nominal Lindblad resonances occur when $D = 0$ such that $\Tilde{\omega}=m(\omega_{ml}-\Omega) = \pm \kappa$, see equation (\ref{eqn:resonance_offset}). In other words the Doppler shifted forcing frequency matches the fluid epicyclic frequency. In a disc with uniform pressure (i.e. $q=p=0$) in which $\Omega = \kappa =\Omega_\textrm{k}$, this corresponds to the standard mean motion resonant locations
\begin{equation}
    \label{eqn:lindblad_resonances}
    \frac{r_{\pm}}{a} = \left(\frac{m\pm 1}{l}\right)^{2/3}.
\end{equation}
Lindblad resonances are the locations where the inertial-acoustic waves are launched, which then propagate away from the planet. For suitably low values of $m \ll r/H$, \cite{GoldreichTremaine_1979} used the WKB approximation to describe the wave launching as occurring strictly at these resonances, although \citet{RafikovPetrovich_2012} have shown that this simple picture is incomplete, which manifests itself in the \textit{negative torque density} phenomenon. 

However, for $m \gtrsim r/H$ the highly non-axisymmetric wave structure modifies the underlying dispersion relation wherein the azimuthal pressure gradients become significant. This causes the effective wave launching resonant positions to be displaced from their fiducial Lindblad value and instead pile up at a  distance of $\sim 2H/3$ from the planet \citep{Artymowicz_1993}. This coincides with the location at which the azimuthal disc shear flow becomes supersonic relative to the planet \citep{GoodmanRafikov_2001} and accounts for the \textit{torque cut-off} phenomenon, wherein the torque contribution from the high-$m$ potential harmonics becomes exponentially suppressed. This deviation of the wave launching from the classical \cite{GoldreichTremaine_1979} WKB approach can lead to notable discrepancies with reality when the higher-$m$ modes are important.   

\subsubsection{Corotation resonances}
\label{subsubsec:corotation}

Another important resonant location occurs where $\tilde{\omega} = 0$ such that $\omega_{ml} = \Omega$. In other words the pattern speed matches the local orbital velocity and they move together, lending to the name \textit{corotation} resonances. For a circular planetary orbit, with only one pattern frequency, the corotation resonance coincides with the coorbital radius to $\mathcal{O}(h_\textrm{p}^2)$. However, eccentric orbits are decomposed into Fourier modes exhibiting a variety of pattern speeds, corresponding to a host of non-coorbital corotation resonances \citep{GoldreichTremaine_1980} which also need to be considered. 

Such corotation resonances are complicated locations in the planet-disc interaction problem which are interpreted in a variety of different regimes \citep[e.g.][]{GoldreichTremaine_1979, Ward_1991, PaardekooperPapaloizou_2009}. Since the methodology in this paper is an entirely linear analysis, we are limited to probing the linear corotation torque as originally investigated by \cite{GoldreichTremaine_1979}. In this regime, the corotation resonance manifests itself via a discontinuity in angular momentum flux across the resonant location (see Section \ref{subsection:circular_wake_amf}), equal to the torque imposed in the vicinity of corotation. The angular momentum flux excited in this region decays evanescently from corotation, indicating that the angular momentum is deposited locally and cannot escape from the vicinity of the resonance; this is in stark contrast to the wavelike propagation of spiral waves away from Lindblad resonances. If efficient communication of this deposited angular momentum is maintained through viscous or diffusive effects, the linear corotation torque \citep{PaardekooperEtAl_2011} is preserved. Otherwise, nonlinearities will develop and manifest as the \textit{horseshoe drag} wherein advective momentum terms govern the exchange of angular momentum as fluid parcels librate around corotation \citep[e.g.][]{Ward_1991,PaardekooperEtAl_2010}. Whilst we will not resolve this nonlinear dynamics in this study, we find that the linear corotation torque is often subdominant compared with the Lindblad torque, see Section \ref{sec:eccentric_torques}. Furthermore, we exploit a standard torque decomposition technique to separate the Lindblad and corotation contributions \citep[see e.g.][]{MirandaRafikov_2020}, which can then be dealt with separately if desired.

\subsection{Numerical methodology}
\label{subsection:numerical}

\subsubsection{Summary}

In order to compute the net wake structure, and associated torque back-reaction onto the planet, we solve equation \eqref{eq:master_eqn} numerically. This immediately allows us to circumvent the compromising assumptions entailed with analytical approaches. In particular, the classic torque formula of \cite{GoldreichTremaine_1979} breaks down when azimuthal pressure gradients enter the problem for large $m\sim h_{\textrm{p}}^{-1}$. By solving for the global structure of the modes directly, we fully capture the excitation of the density wave and sample its interaction with the perturbing potential over the finite excitation interval, rather than just at isolated resonances. Motivated by this we can readily exploit the numerical method tested and used in \paper1. This method implements a solution technique largely inspired by the work of \cite{KorycanskyPollack_1993} and developed for circular planet-disc interactions by \cite{MirandaRafikov_2020}. For full details we refer the reader to these works and only summarise the key steps here. 

Firstly, the orbit of the planet is specified by a numerical solution to Kepler's equation. Since we are interested in probing larger values of the eccentricity, we cannot simply use the leading order epicylic approximation which expands the potential perturbation in terms of Laplace coefficients \citep[e.g.][]{GoldreichTremaine_1980}. Indeed, as eccentricity increases, this expansion converges slowly and becomes inaccurate. Instead, the perturbing potential is defined across the disc at each time during the planetary orbit through equation \eqref{eqn:direct_potential}. By direct numerical integration across the temporal and azimuthal periods, we numerically extract the Fourier coefficients $\Phi_{ml}(r)$ at each radius in the disc according to
\begin{align}
    \label{eq:phi_mode_integral}
    \Phi_{ml}(r) &= \frac{1}{2\pi^2}\int_{t = 0}^{2\pi/\Omega_\textrm{p}} \int_{\phi = 0}^{2\pi} \Phi(r,\phi,t) \cos(m\phi-l \Omega_\textrm{p} t) \,d\phi\,dt \nonumber\\
             &= -\frac{GM_{\text{p}}}{2\pi^2}\int_{t = 0}^{2\pi/\Omega_\textrm{p}} \int_{\phi = 0}^{2\pi} \frac{\cos(m\phi-l \Omega_\textrm{p} t) \,d\phi\,dt}{\left[r^2+r_{\text{p}}^2-2 r r_{\text{p}} \cos\psi+\epsilon^2\right]^{1/2}} . 
\end{align}
Here $\psi(t) = \phi-\phi_{\text{p}}(t)$ denotes the azimuthal displacement from the planet and $\epsilon(|{\bf r}-{\bf r}_\textrm{p}|)$ is the softening kernel, employed to prevent singularities at the location of the planet (when $r - r_{\text{p}} = \psi = 0$) and to mimic the vertical averaging of the three dimensional disc in this 2D setup. Throughout this work we will use the simplest Plummer softening prescription $\epsilon = b H(r_\textrm{p})$, for a choice of constant parameter $b$. Recent work usually assumes the softening to be a significant fraction of the local scale height $H_\textrm{p}$, i.e. $b\sim \mathcal{O}(1)$, so as to simulate the effects of a 3D potential \citep[e.g.][]{PaardekooperPapaloizou_2009}. 

This method for numerical extraction of potential harmonics $\Phi_{ml}$ has proven successful in our previous synthesis of wake morphologies in \paper1. Note that in equation \eqref{eq:phi_mode_integral} we are neglecting the indirect potential component but do compute the direct $m=1$ term. Inspection of equation \eqref{eqn:lindblad_resonances} shows that the nominal inner Lindblad resonance vanishes for this Fourier component, which makes the inner boundary condition numerically unstable and a higher order scheme must be used to solve for the wave. Whilst this $m=1$ term does not change the qualitative wake morphology, we find that it does provide a non-negligible contribution to the torque quantities and should be included (see Fig.~\ref{fig:circular_torques} and discussion in Section \ref{subsection:circular_theory}).

The potential harmonics $\Phi_{ml}$ are evaluated over a discrete radial grid and then fed into the master equation \eqref{eq:master_eqn} for a given choice of $m$ and $l$. The potential at arbitrary values of $r$ can then be interpolated as required. A pair of homogeneous, unforced solutions as well as the forced, inhomogeneous response are integrated outwards from the mode-specific corotation radius where $\omega_{ml}=\Omega$ towards the inner and outer boundaries at $r/a = 0.05$ and $5.0$ respectively. The relative contributions from the unforced wave components are fixed by specifying outgoing radiative boundary conditions at the inner and outer disc edges. Spurious incoming waves are removed by adjusting the boundary conditions using a phase-gradient error minimisation technique \citep{KorycanskyPollack_1993}. The net wake is finally constructed as a superposition of all these individual modes.

\subsubsection{Convergence}
\label{subsubsection:convergence}

Qualitatively, integers $m$ and $l$ govern the azimuthal and temporal convergence of our solutions, respectively. For a circular planet, the wake structure is stationary in a frame which corotates with the orbit, and hence $l=m$, i.e. the pattern speed of modes is equal to the mean motion $\Omega_\mathrm{p}$. However, as planetary eccentricity becomes non-zero, epicylic motion of the planet about the guiding centre introduces an essential time-dependence and `off-diagonal' $|l-m|\neq 0$ modes begin to contribute to forcing. Indeed, the classic disturbing Function for a perturbing planet with small $e$ has a modal expansion which goes as $\Phi_{lm}\propto e^{|l-m|}$ \citep[see][]{MurrayDermott_1999}. Therefore, as we boost $e$ to larger values, we need to include modes of higher order in $|l-m|$ in our analysis. Furthermore, whilst the qualitative morphology of the wake structure might appear to converge rather quickly, special care should also be taken to ensure that the second order quantities such as the torque and angular momentum flux have converged as well when $m$ and $|l-m|$ increase. This ever more challenging convergence criterion with increasing eccentricity ultimately limits our ability to probe arbitrarily high values of $e$. We therefore  restrict our attention to cases where we are confident of convergence, as discussed in more detail in Appendix \ref{app:torque_converegence}. Furthermore, during the post-processing of the many thousands of calculated modes, we very occasionally find some (numerically) spurious behaviour in a few $m=1$ waves which would produce slight deviations in the broad trends discussed later in this paper. We discuss this anomalous behaviour in Appendix \ref{app:m1q1}. Ultimately, these modes are filtered out in our final results, as justified by the compatibility of the adjusted results with adjacent points in parameter space.

\subsubsection{Computing the torque and angular momentum flux}
\label{subsubsection:torque_computation}

With the wave modes in hand, we are in a position to construct second order quantities such as the torque density and angular momentum flux (AMF). These are second order in that they involve a product of linear perturbations. Indeed, the AMF is defined as 
\begin{equation}
    F_{J}(r)  = r^2 \Sigma(r) \oint \operatorname{Re}[\delta u_r(r,\phi)] \operatorname{Re}[\delta u_\phi (r,\phi)] d\phi ,
    \label{eq:AMF-def}
\end{equation}
where the net velocity perturbations are given by a sum over modes
\begin{equation}
    \delta u_r(r,\phi) = \sum_{m,l} \delta u_{r,ml}(r) \exp{[\mathrm{i}(m\phi-l\Omega_\textrm{p} t)]},
\end{equation}
\begin{equation}
    \delta u_\phi(r,\phi) = \sum_{m^\prime,l^\prime} \delta u_{\phi,m^\prime l^\prime}(r) \exp{[\mathrm{i}(m^\prime\phi-l^\prime\Omega_\textrm{p} t)]}.
\end{equation}
Inserting these expressions into the equation (\ref{eq:AMF-def}) and simplifying we see that 
\begin{align}
    F_{J}(r)  = \pi r^2 \Sigma(r) \sum_{m} & \sum_{l,l^\prime} \operatorname{Re}[\delta u_{r, ml} \delta u_{\phi, ml^\prime}^{*}] \cos[(l-l^\prime)\Omega_\textrm{p}t]\nonumber\\
    & +\operatorname{Im}[\delta u_{r, ml} \delta u_{\phi, ml^\prime}^{*}] \sin[(l-l^\prime)\Omega_\textrm{p}t],
\end{align}
where $^{*}$ denotes the complex conjugate. Similarly the radial torque density is defined as
\begin{equation}
    \frac{dT}{dr} = -r \oint \operatorname{Re}[\delta\Sigma(r,\phi)]\operatorname{Re}\left[\frac{\partial \Phi_\textrm{p}}{\partial\phi}\right]d\phi ,
\end{equation}
which can be written in terms of the modal contributions as
\begin{align}
\frac{dT}{dr} = -\pi r \sum_{m} \sum_{l,l^\prime} m & \left\{ \cos[(l-l^\prime)\Omega_\textrm{p}t] \Phi_{ml^\prime} \operatorname{Im} [\delta \Sigma_{ml}] \right. 
\nonumber\\
& \left. -\sin[(l-l^\prime)\Omega_\textrm{p}t] \Phi_{ml^\prime} \operatorname{Re} [\delta \Sigma_{ml}]\right\}.
\label{eq:time_dependent_dTdr}
\end{align}
Notice that the azimuthally-averaged quantities fix $m=m^\prime$ so only equal azimuthal wavenumbers couple together and contribute to the orbit-averaged torque. Furthermore, the temporal dependence enters through the sinusoidal argument $(l-l^\prime)$ and so the AMF and torque density will oscillate over the orbital period (see eccentric results in Section \ref{subsection:radial_torque_structure} and Fig.~\ref{fig:time_dependent_torque}). We are primarily interested in the cumulative, secular effects over many orbital periods and thus average over the mean motion period giving us
\begin{align}
    & \langle F_{J}(r) \rangle = \sum_{m,l} F_{J, ml}, \\
    &\left\langle \frac{dT}{dr} \right\rangle = \sum_{m,l} \frac{dT_{ml}}{dr} ,
\end{align}
where $\langle\cdot\rangle = (\Omega_\textrm{p}/2\pi)\oint\cdot dt$ and we define the individual modal contributions to be
\begin{align}
    \label{eq:F_jml}
    F_{J, ml} & = \pi r^2 \Sigma(r) \operatorname{Re}[\delta u_{r, ml} \delta u_{\phi, ml}^{*}], \\
    \label{eq:dTdr_jml}
    \frac{dT_{ml}}{dr} &= -\pi r m \Phi_{ml} \operatorname{Im} [\delta \Sigma_{ml}].
\end{align}
Henceforth we shall drop the angled brackets and simply remember that we are dealing with the averaged torque and AMF quantities in the remainder of this paper, unless stated otherwise. We also define the time-averaged, radially-integrated torque to be
\begin{equation}
    \label{eq:one_sided_torque}
    T(r) = \int_a^{r} \frac{dT}{dr} dr.
\end{equation}
Although this one-sided integral differs compared with some previous conventions, it has been used recently by \cite{CimermanEtAl_2024} to clearly partition the torque contributions interior and exterior to the planet. Indeed with this definition we are integrating outwards from the guiding centre radius so that the net torque acting on the entire disc is given by the difference (rather than the sum) $T_\textrm{net} = T(r_\textrm{out})-T(r_\textrm{in})$. Here, $r_\textrm{in}$ and $r_\textrm{out}$ denote the inner and outer disc boundaries respectively. For example, a positive value of $T(r>a) > 0$ arises from positive torque density contributions depositing angular momentum in the outer disc. The equal and opposite back reaction on the perturber then extracts planetary angular momentum and (for a circular orbit) promotes inwards migration. Meanwhile a positive value of $T(r<a) > 0$ owes to a negative torque density which extracts angular momentum from the inner disc. This is accordingly injected into the planet and (for a circular orbit) drives outwards migration. This sign convention is chosen to allow us to easily compare the asymmetry of the inner and outer disc torques (as seen in Fig.~\ref{fig:circular_morphology_torques} and \ref{fig:torque_density_integrated_torque}) -- highlighting the `repulsive' competition either side of the guiding centre radius.

It is also useful to decompose this net torque into the separate Lindblad and corotation contributions using a recipe similar to that employed by \cite{MirandaRafikov_2020}. By linearity, the total torque acting over the full disc can be calculated for each separate modal contribution
\begin{equation}
    T_{\textrm{net},ml} = \int_{r_\textrm{in}}^{r_\textrm{out}} \frac{dT_{ml}}{dr}dr = T_{ml}(r_\textrm{out})-T_{ml}(r_\textrm{in}),
\end{equation}
which are then summed to yield the net torque $T_\textrm{net} = \sum_{m,l} T_{\textrm{net},ml}$. At the inner and outer boundaries, sufficiently far from the planet, the torque contributions have plateaued (see discussion in Section \ref{subsection:circular_wake_amf} and the black lines in the associated Fig.~\ref{fig:circular_morphology_torques}). 

The corotation torque is now extracted from every $T_{ml}$ modal combination by looking for the AMF discontinuity in the vicinity of each $\omega_{ml} = \Omega$ resonance. Since we only want to extract the discontinuity about a specific point, we must precisely pinpoint the location of the corotation resonance. To second order in the aspect ratio $h_\mathrm{p}$, the corotation resonance occurs at
\begin{equation}
    r_{\textrm{C},ml} = a \tilde{r}_{\textrm{C},ml}\left[1-\frac{1}{3} h_\textrm{p}^2 (q+p) \tilde{r}_{\textrm{C},ml}^{1-q}\right],
\end{equation}
where $\tilde{r}_{\textrm{C},ml} = (l/m)^{-2/3}$ is the nominal corotation resonance location in a pressureless disc normalised by $a$, for a given $(m,l)$ mode. We then numerically interpolate the calculated $F_{J,ml}(r)$ and $T_{ml}(r)$, which are evaluated at discrete grid locations, onto points on either side of $r_{\textrm{C},ml}$ such that 
\begin{align}
    &\Delta F_{J,ml} = F_{J,ml}(r_{\textrm{C},ml}+\Delta r) - F_{J,ml}(r_{\textrm{C},ml}-\Delta r),\\
    &\Delta T_{ml} = T_{ml}(r_{\textrm{C},ml}+\Delta r) - T_{ml}(r_{\textrm{C},ml}-\Delta r),
\end{align}
where the narrow width $\Delta r = 0.005a$ is chosen to encompass the AMF discontinuity. The modal corotation torque acting on the disc is then computed via the difference
\begin{equation}
    T_{\textrm{C},ml} = \Delta T_{ml}-\Delta F_{J,ml},
\end{equation}
where the inclusion of the torque differential here offsets the change in AMF due to $dT_{ml}/dr$ over this small but finite interval. Finally, the remaining modal Lindblad torque is computed as
\begin{equation}
    T_{\textrm{L},ml} = T_{\textrm{net},ml}-T_{\textrm{C},ml}.
\end{equation}
This decomposition process is repeated separately for each modal combination before the individual results are finally summed to obtain the partitioning of the net torque into the total corotation and Lindblad contributions; $T_\textrm{C} = \sum_{m,l} T_{\textrm{C},ml}$ and $T_{L} = \sum_{m,l} T_{\textrm{L},ml}$, respectively.

%% file: Sections/3_circular_benchmark.tex
\section{Circular benchmark}
\label{sec:circular_benchmark}

We start presentation of our results by demonstrating the performance of our method for the case of a circular planetary orbit with $e=0$. We benchmark our framework against many standard planet-disc interaction results, including \cite{KorycanskyPollack_1993} who used an approach identical to ours in a circular case, and present a useful point of comparison as we generalise to eccentric orbits.

\subsection{Numerical parameters}
\label{subsec:numerical_params}

To compute angular momentum characteristics of circular planet-disc coupling, we sum contributions of modes from $m_\textrm{min} = 1$ to $m_\textrm{max} = 150$. Furthermore, since $e=0$ we only need to consider the diagonal temporal contributions $l=m$. We adopt the softening parameter $b=0.3$ and consider the wake structure across a grid of typical disc power-law structures: the surface density exponent $p\in [0.0,0.5,1.0,1.5]$, the temperature exponent $q\in [0.0,0.25,0.5,0.75]$ and the disc aspect ratios $h_\textrm{p}\in[0.06,0.07,0.08,0.09,0.10]$. This range spans from uniform surface densities, to steeper profiles more typically inferred from observations. Furthermore, the temperature profile ranges from being a somewhat artificial, globally isothermal case, to a maximal temperature gradient predicted by the `lamp post' illumination model of passive disc heating \citep[see][]{Friedjung_1985}.

\subsection{Qualitative Wake and AMF Features}
\label{subsection:circular_wake_amf}

\begin{figure*}
    \centering
    \includegraphics[width=\textwidth]{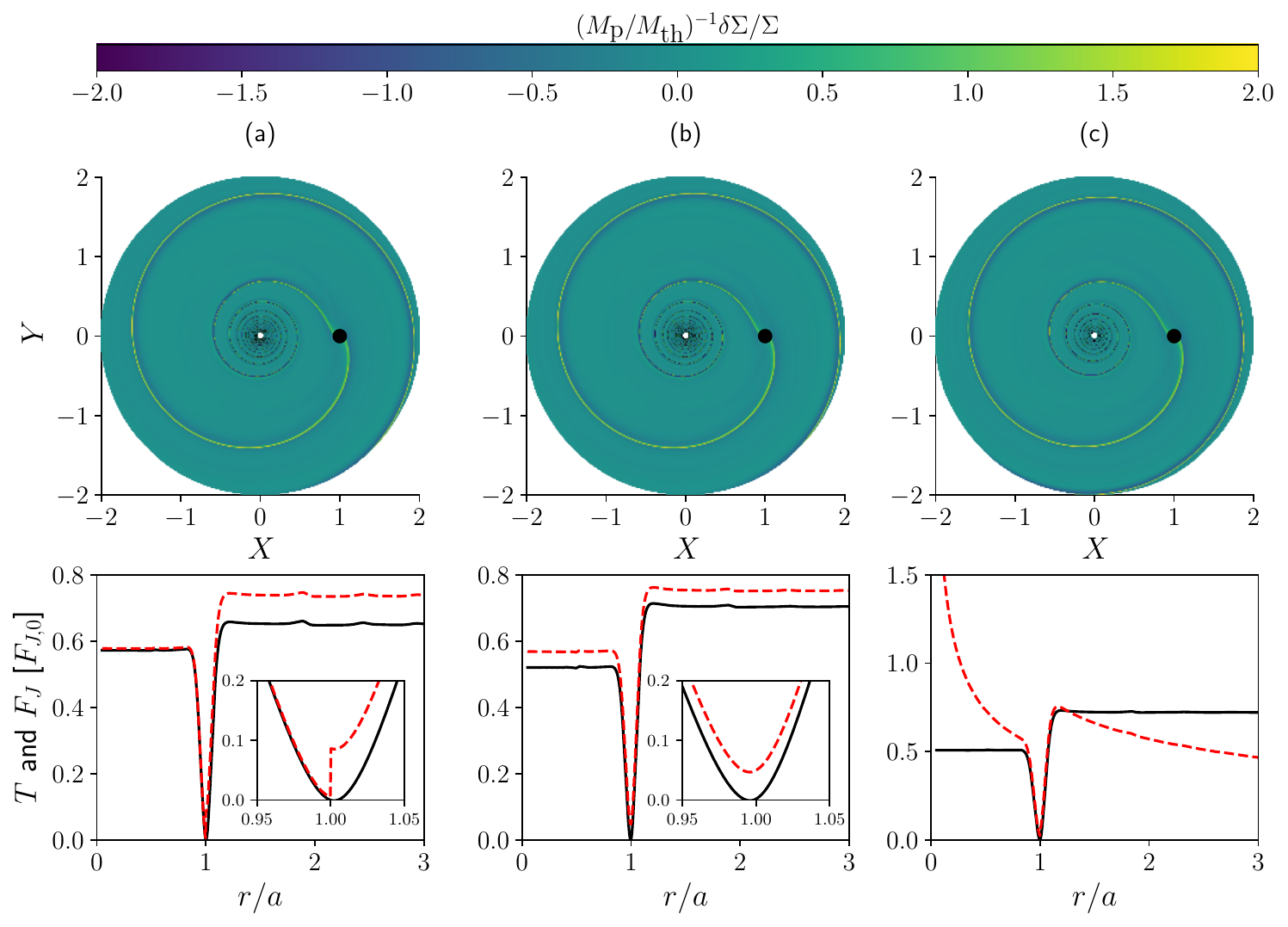}
    \caption{\textit{Upper panels}: The relative surface density perturbation for three different choices of $(q,p)$. From left to right (a) $= (0.0,0.0)$, (b) $= (0.0,1.5)$ and (c) $ = (0.5,1.5)$. \textit{Lower panels}: the corresponding radial torque profiles $T(r)$ (black lines) and the angular momentum flux $F_{J}(r)$ (red dashed lines) in characteristic flux units $F_{J,0}$ defined by the equation (\ref{eq:FJ0}). The inset panels zoom in to the coorbital region.}
    \label{fig:circular_morphology_torques}
\end{figure*}

The upper row of Fig.~\ref{fig:circular_morphology_torques} shows three example wake morphologies. Here we have specifically chosen $(q,p) = \lbrace(0.0,0.0), (0.0,1.5) , (0.5,1.5)\rbrace$ moving from left to right so as to exemplify different features in the excited AMF response ($h_\textrm{p} = 0.06$ in all cases). Notice that the relative surface density perturbation morphology for the circular planet is dominated by a single, constructive spiral arm as understood by \cite{OgilvieLubow_2003} and \citet{Rafikov2002a}, although this picture breaks down in the inner disc where multiple spiral arms emerge \citep{Bae2018,MirandaRafikov_2019a}. Only subtle qualitative differences are seen in the winding of the arm as the temperature exponent changes. 

Meanwhile more obvious differences are apparent in the corresponding integrated radial torque and AMF profiles shown in the second row. The solid black and red dashed lines plot $T(r)$ and $F_{J}(r)$, respectively, in characteristic flux units \citep{GoldreichTremaine_1980}
\begin{equation}
    F_{J,0} \equiv \Sigma_\textrm{p} a^4 \Omega_\textrm{p}^2 h_\textrm{p}^{-3} (M_\textrm{p}/M_*)^2.
    \label{eq:FJ0}
\end{equation}

In column (a) we consider a globally isothermal, uniform surface density disc with $q=0.0$ and $p=0.0$. We see that the net torque rises steeply away from the coorbital location as the dominant Lindblad resonances contribute to $T(r)$ within a distance of $\mathcal{O}(H)$ from the planet. Away from this region the torque levels out, although in the outer disc one may also notice small `wiggles' in both $T(r)$ and $F_J(r)$, the nature of which has recently been understood by \cite{CimermanEtAl_2024}. The difference between the plateaued torques at the inner and outer boundaries therefore provides the net torque acting back on the planet. The AMF profile exhibits a similar behaviour up to some constant offset in the outer disc region. The conservation of AMF at large distances from the planet is expected since $F_{J}$ is the wave-action for waves propagating through a globally isothermal disc \citep[see][]{MirandaRafikov_2020}. The vertical offset of $F_{J}$ from $T$ in the outer disc can be traced back to the torque exerted at the corotation resonance where there is a discontinuity in the AMF. Zooming into this region within the inset panel shows a step-wise jump whose magnitude gives the linear corotation torque (see Section \ref{subsubsec:corotation}).

In column (b) we again take a globally isothermal disc with $q = 0.0$ but this time adopt a surface density profile with $p = 1.5$. This special choice of surface density ensures that the radial vortensity gradient of the background disc is zero. Indeed, for an axisymmetric disc, the vertical component of vortensity (normal to the disc plane) is 
\begin{align}
\label{eq:vortensity}
    \varpi &\equiv \frac{1}{r\Sigma}\frac{\partial}{\partial r}\left(r^2 \Omega\right) \nonumber \\
    &\approx \frac{\Omega_\textrm{p}}{2\Sigma_\textrm{p}}\left[ \left(\frac{r}{a}\right)^{p-3/2}-h_\textrm{p}^2(q+p)(3/2-q)\left(\frac{r}{a}\right)^{p-q-1/2}\right],
\end{align}
to second order in the aspect ratio $h_\textrm{p}$. Thus, for $p=3/2$ the dominant, leading order contribution (first term) to $\varpi$ is constant with $r$. For a barotropic disc, \cite{GoldreichTremaine_1979} showed that the linear corotation torque is proportional to $\partial(\varpi^{-1})/\partial r$ evaluated at corotation; it should thus vanish for $p=3/2$ to $\mathcal{O}(h_\textrm{p})$. As a result, the zoomed inset panel in Fig.~\ref{fig:circular_morphology_torques}(b) shows no evidence of a discontinuity in the red dashed line, which is obvious in Fig.~\ref{fig:circular_morphology_torques}(a). However, there is a uniform offset between the integrated torque $T(r)$ and angular momentum flux $F_J(r)$ as previously seen by \cite{KorycanskyPollack_1993}, \cite{DongEtAl_2011} and \cite{RafikovPetrovich_2012}. Whilst the physical origin of this constant offset is not yet understood, the conservation of angular momentum still holds since $dT/dr = dF_{J}/dr$ everywhere. 

In column (c) we again take $p=1.5$ to minimize the effect of the corotation resonance. However, now we adopt a locally isothermal EoS with $q = 0.5$. In this case the torque and AMF profiles clearly do not overlap (i.e. they are not related by a constant offset) and $dT/dr \neq dF_{J}/dr$. This behaviour was noted for locally isothermal EoS by \cite{MirandaRafikov_2020} who examined the excitation and propagation of density waves for different thermal cooling prescriptions. They also showed that in the locally isothermal regime the conserved wave action is given by $F_{J}/c_s^2$ instead of $F_J$. This wave action obviously reduces to the aforementioned globally isothermal result when $c_s$ is radially constant. Keeping this phenomenology in mind will be important as we extend our results to the eccentric interaction.

\subsection{Comparison with the existing literature}
\label{subsection:circular_theory}

Our direct numerical method circumvents the various approximations used in the past in producing analytical estimates of the torque. Here we briefly compare our methodology with previous torque prescriptions to emphasize the benefits of our approach.

We decompose the net torque into the Lindblad and corotation components in accordance with the method described in Section \ref{subsubsection:torque_computation}. Note that since we are now focusing on the case of circular planetary orbits, where only the $m=l$ modes contribute, all corotation resonances occur at the same location near $r_\mathrm{p}$ and can be extracted in one go from the single jump in the angular momentum flux profile, see Figure \ref{fig:circular_morphology_torques}(a). This is a simplification to the general pipeline where the corotation discontinuities occur at different radial locations and must be extracted separately for each modal combination before being summed together (as detailed in Section \ref{subsection:non-coorbital resonances}).

\begin{figure}
    \centering
    \includegraphics[width=0.9\columnwidth]{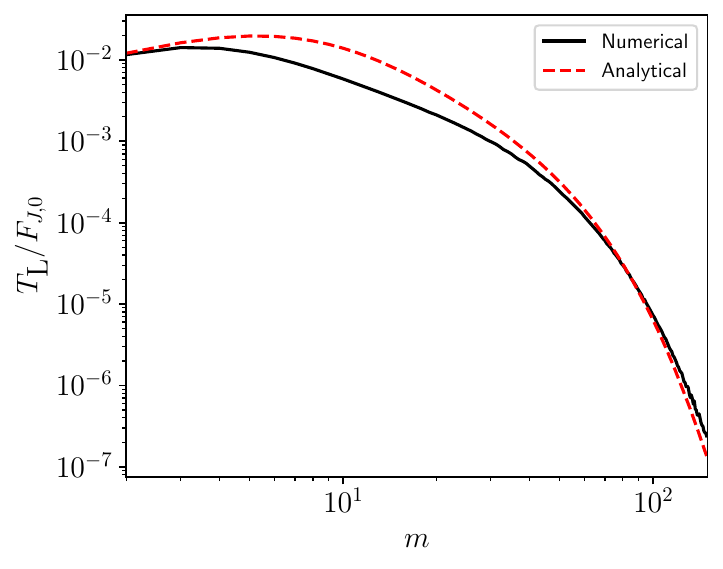}
    \caption{Comparison of the torque contributions as a function of the modal number $m$ for a disc with $(q,p,h,b) = (0.0,0.0,0.06,0.3)$. \textit{Red dashed line}: theoretical net Lindblad torque from equation \eqref{eq:L_WKB}. \textit{Solid black line}: numerically extracted Lindblad torque as described in the text.}
    \label{fig:analytical_theory_torque}
\end{figure}

In Fig.~\ref{fig:analytical_theory_torque} we show the contribution of individual modes to the net Lindblad torque for the case of $p=0.0$, $q=0.0$, $h_\mathrm{p}=0.06$ and $b=0.3$. Our results are shown as the solid black line. Notice that the dominant contributions occur for relatively small values of $m$, before experiencing the torque cut-off at larger values of $m$. We compare this numerical result with the modified WKB analytical prescription for Lindblad torques acting on the disc (dashed red curve) given by equation (26) of \cite{PapaloizouLarwood_2000},
\begin{align}
    \label{eq:L_WKB}
    T_{\textrm{L},\textrm{WKB}} = & \frac{S \pi^2 \Sigma}{3 \Omega \Omega_\textrm{p}} 
    \left[r\frac{d\Phi_{m}}{dr}+\frac{2 m^2 (\Omega-\Omega_\textrm{p})\Phi_{m}}{\Omega}\right]^2 \nonumber \\
    & \times \left\lbrace \frac{\left[1+m^2 c^2/(r\Omega)^2\right]^{-1/2}}{1+4m^2c^2/(r\Omega)^2}\right\rbrace ,
\end{align}
which is to be evaluated at the inner and outer Lindblad resonance locations where $S=-1$ and $+1$ respectively. The first part of equation \eqref{eq:L_WKB} simply corresponds to the classic result found in \cite{GoldreichTremaine_1979} whilst the multiplicative modification in the second line accounts for the pressure-induced correction introduced by \cite{Artymowicz_1993}, relevant for $m \gtrsim h_\textrm{p}^{-1}$. Whilst the two curves qualitatively agree, there are clear discrepancies within the range $m\sim 5-50$ which dominates the total torque. This echoes the findings of \cite{KorycanskyPollack_1993} who also plot the deviation between their method and analytical predictions when adopting similar disc properties. This difference highlights the benefits of using our global approach which avoid the limiting assumptions of purely analytical methods. 

It should be noted that the comparison between our results and theoretical predictions in Figure \ref{fig:analytical_theory_torque} appears qualitatively different from the similar comparison conducted by \cite{KorycanskyPollack_1993}, see their Figure 9. This might be due to the differences in the potential softening used and/or the different analytical predictions being used for comparison. For example, whilst we plot the \cite{Artymowicz_1993} modified WKB formulae at the shifted Lindblad resonances as suggested by \cite{PapaloizouLarwood_2000}, \cite{KorycanskyPollack_1993} compare their numerical results with both the cut-off modified prescription evaluated at the nominal Lindblad resonances and the unmodified \cite{GoldreichTremaine_1979} WKB expression evaluated at shifted resonance positions. 

\begin{figure*}
    \centering
    \includegraphics[width=\textwidth]{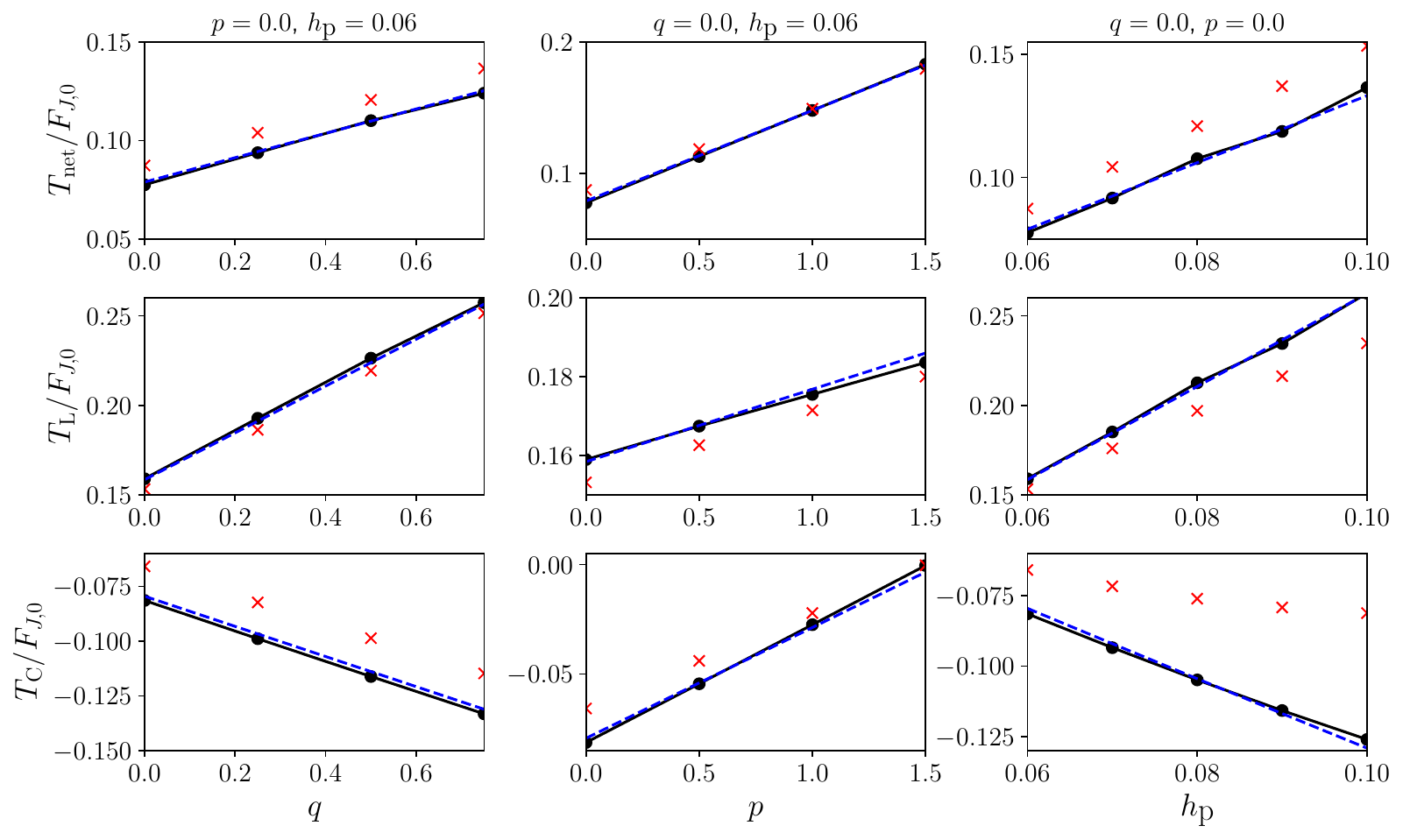}
    \caption{The back-reaction torque (in characteristic flux units $F_{J,0}$) of the excited density wake onto the circularly orbiting planet. \textit{Upper row}: net integrated torque $T_\textrm{net}$. \textit{Middle row}: extracted Lindblad torque $T_\textrm{L}$. \textit{Lower row}: extracted corotation torque $T_C$. The columns from left to right show the variation in these quantities with disc parameters $q$, $p$ and $h_\textrm{p}$ respectively. The red crosses show the data points computed neglecting the $m=1$ mode. The black points and connective lines show the data points computed including the $m=1$ mode. The dashed blue lines show the best linear fits (\ref{eq:Tnet_fit})-(\ref{eq:TL_fit}) to the full data cube.}
    \label{fig:circular_torques}
\end{figure*}

We also explore how the full (integrated over $m$) circular torques vary with the disc properties. We demonstrate the total, Lindblad and corotation torques in Fig.~\ref{fig:circular_torques} along particular slices through the full $(q,p,h_\textrm{p})$ parameter space (outlined in Section \ref{subsec:numerical_params}). 
The upper, middle and lower rows plot the values of $T_\textrm{net}$, $T_\textrm{L}$ and $T_\textrm{C}$ respectively. Meanwhile, the first column shows the variation in torques with $q$ whilst $p=0.0$ and $h_\textrm{p}=0.06$ are held constant. Similarly, the middle column fixes $q=0.0$ and $h_\textrm{p}=0.06$ whilst p varies. Finally the last column fixes $q = 0.0$ and $p=0.0$ whilst $h_\textrm{p}$ varies. The qualitative behaviour in these slices is illustrative of the wider parameter space and the full data cube is available in the supplementary data (see Appendix \ref{app:parameter_space}). 

The black dots and red crosses correspond to the torque values computed including and excluding the $m=1$ mode, respectively. Clearly, these values are quite different, meaning that the $m=1$ mode provides a non-negligible contribution to the torques. This might be anticipated for disc aspect ratios which are moderately thick where the torque cut-off quenches the higher order modes and therefore increases the relative importance of the lower modes. In most of the points plotted we see that the inclusion of the $m=1$ mode slightly reduces the net torque. This is because the enhancement of the negative corotation torque contribution caused by including $m=1$ mode tends to outweigh the slight increase of the Lindblad torque coming from the additional outer $m=1$ Lindblad resonance. 

It is also clear that all these torques are well fit by straight lines. More generally we perform a least squares minimisation of the three dimensional hyperplane fitting function in order to quantitatively extract these trends. We find the following fits 
\begin{align}
    & T_\textrm{net} = (1.413+1.109 q+1.237 p-0.017 q p)h_\textrm{p}^{1.025} F_{J,0} 
    \label{eq:Tnet_fit}\\
    & T_\textrm{L} = (2.557+2.113 q + 0.297 p -0.034 q p)h_\textrm{p}^{0.989} F_{J,0}
    \label{eq:TL_fit}
\end{align}
which are over-plotted as the blue dashed lines. We acknowledge that the ratio of the coefficients prepending $q$ and $p$ are notably different from those found by \cite{KorycanskyPollack_1993} in equations (32). This can be attributed to the different softening lengths being used. Whilst we adopt a softening parameter of $b = 0.3$, they use a much smaller value of $0.003$. Such a small softening leads to a sharper perturbing potential structure which cannot be easily resolved without extending the calculation to much higher values of $m$. This becomes computationally expensive for our generalised method, particularly for the eccentric case where we also have to compute off-diagonal $l \neq m$ modes. On the other hand, the larger softening length chosen here is more consistent with 2D studies of planet-disc interaction \citep[e.g.][]{CresswellNelson_2006, PaardekooperEtAl_2010} and hence is suitable for our purposes. Nonetheless, this sensitivity to parameters emphasizes the need to be careful when interpreting quantitative statements. This point is also raised by \cite{PaardekooperPapaloizou_2009} and \cite{PaardekooperEtAl_2010} who discuss that the linear dependencies on $q$ and $p$ will change depending on the choice of softening parameter $b$ and on whether the calculation is 2D or 3D. Indeed, they comment that the simple power law scaling in $b$, often tacked on to torque prescriptions, is unreliable and thus we caution against blindly using this to adjust for different softening values. As for all previous 2D planet-disc studies, our results should be regarded in the context of the region of parameter space explored. In the future we would like to extend our parameter sweep to a range of values in $b$ to investigate this effect further.

%% file: Sections/4_eccentric_torques.tex
\section{Torque Results for Eccentric Planets}
\label{sec:eccentric_torques}

\subsection{Parameter space and numerical considerations}
\label{subsection:eccentric_parameter_space}

With our machinery tested on circular orbits, we are now prepared to turn our attention to the case of eccentric planet-disc interaction. Once again, we are interested to see how the torques change as we vary the disc temperature and surface density profiles and so we adopt $q \in [0.0,0.25,0.5,0.75]$ and $p \in [0.0,0.5,1.0,1.5]$. We also vary the disc aspect ratio such that $h_\textrm{p} \in [0.06,0.08,0.1]$, although for much of the discussion we will focus on the results specific to $h_\textrm{p} = 0.06$. Throughout our main parameter study we maintain a softening coefficient of $b=0.3$, although we acknowledge this free choice is one of the main caveats of the 2D formalism.

The relevant quantity which governs the nature of the interaction is the ratio $\tilde{e} \equiv e/h_\textrm{p}$. Indeed as $\tilde{e}$ exceeds unity, the planet will undergo transonic motion relative to the background gas, marking a distinct change in the wave excitation behaviour (see Section \ref{subsection:radial_torque_structure}). Thus we are particularly interested in probing this transition and as far beyond as possible. This turns out to be difficult for numerical reasons. 

One way to increase $\tilde e$ is to boost planetary $e$, but this inevitably increases $|l-m|$ of the modes that must be included, which is challenging (see convergence behaviour in Section \ref{subsubsection:convergence} and Appendix \ref{app:torque_converegence}). Alternatively, one might increase $\tilde e$ by decreasing $h_\textrm{p}$, thus probing the transonic crossing for lower $e$ and hence requiring a reduced expansion in the off-diagonal elements $|l-m|$. However, there is inevitably a price to pay. As the aspect ratio decreases at fixed $\tilde e\sim 1$, so too must the associated transonic radial epicylic excursions of the planet as the distance between pericenter and apocenter $\sim 2H_\textrm{p}$ in this regime. In order to resolve this region without washing out the planetary potential perturbation, we must maintain the same softening parameter $b$ so the smoothing scale decreases as well. This renders the potential structure harder to resolve and demands more azimuthal modes $m$ to be included. 

By experimenting with this trade-off, we found our lowest adopted value of $h_\textrm{p} = 0.06$ to be the optimal compromise for probing the transonic regime, for which we are able to obtain converged torques for $\tilde{e}$ ranging from $0$ to $2$. Here, the upper value of $\Tilde{e} = 2$ corresponds to the critical case for which the motion of the planet relative to the gas remains supersonic for all phases of the orbital motion.

Across this $(N_q\times N_p \times N_h)=(4 \times 4 \times 3)$ disc parameter space we evaluate the mode structure for $e$ ranging between $0.0$ and $0.12$ in $0.01$ intervals. For $e \leq 0.06$ we find that $m_\textrm{max} = 150$ is sufficient. However, for $e > 0.06$ we include modes up to $m_\textrm{max} = 175$ to ensure convergence. In all cases we include off-diagonal temporal contributions up to $|l-m|_\textrm{max} = 40$. This means that for each net wake structure we need to compute $\mathcal{O}(10^4)$ individual modes. For a more quantitative discussion of convergence, we refer the reader to additional details in Appendix \ref{app:torque_converegence}.

\subsection{Radial Torque Structure}
\label{subsection:radial_torque_structure}

\begin{figure}
    \centering
    \includegraphics[width=\columnwidth]{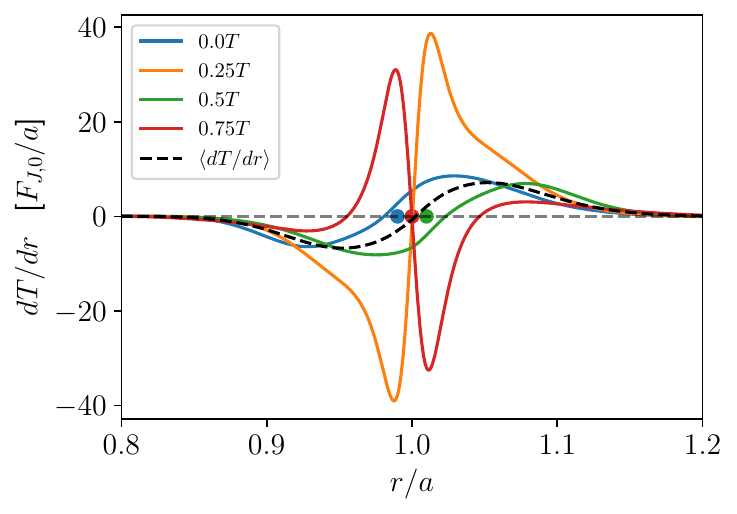}
    \caption{The instantaneous, time-dependent radial torque density at 4 orbital phases of the planetary orbit (noted by the inset labels). The corresponding coloured dots denote the planetary radius at these times. The orbit-averaged torque density is also shown by the black dashed line.}
    \label{fig:time_dependent_torque}
\end{figure}

To gain some intuition, we will first examine the particular case of a  globally isothermal ($q=0.0$) disc, with uniform surface density ($p=0.0$) and an aspect ratio of $h_\textrm{p} = 0.06$. For a non-zero eccentricity, the epicylic motion of the planet introduces a time-dependent wake, so the instantaneous torque also varies over the phase of the planetary orbit as predicted by equation \eqref{eq:time_dependent_dTdr}. This is visualised in Fig.~\ref{fig:time_dependent_torque} for $e = 0.01$, where the instantaneous radial torque density is plotted at 4 phases of the planetary orbit. The blue and green curves denote $dT/dr$ for the pericenter and apocenter times respectively (as indicated by the correspondingly coloured dots indicating the planetary position). Between these times, at quarter periods when the planet is crossing the guiding centre radius $r=a$, the torque density exhibits strong enhancements as the planet chases the exterior (interior) wave launched at pericenter (apocenter). However, for the remainder of this paper we are primarily interested in the epicyclic phase-averaged torque represented by the dashed black curve, which has a modest amplitude indicating that strong  cancellation of instantaneous torques takes place upon their integration  in the course of epicyclic motion. Also note that far from the planet, at $|r-a|\gg ea$, the instantaneous torque density becomes independent of the epicyclic phase, as expected.

\begin{figure*}
    \centering
    \includegraphics[width=\textwidth]{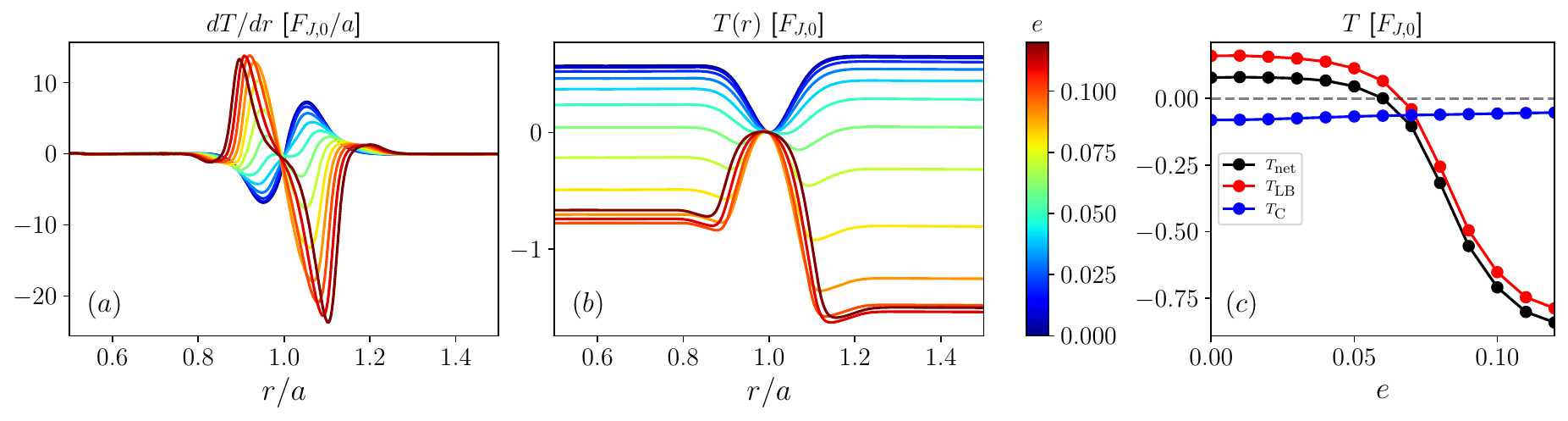}
    \caption{Orbit averaged torque profiles for example case of $p = 0.0$ and $q=0.0$ disc with $h_\textrm{p}=0.06$. \textit{Left panel (a)}: radial torque density $dT/dr$. \textit{Middle panel (b)}: Integrated torque $T$. The different lines correspond to varying eccentricities. The progression from cool to hot colours indicates an increase in $e$ from $0.0$ to $0.12$. \textit{Right panel (c)}: total torque decomposition as a function of eccentricity. The black, red and blue points denote the $T_\textrm{net}$, $T_\textrm{L}$ and $T_\textrm{C}$ respectively.}
    \label{fig:torque_density_integrated_torque}
\end{figure*}

In Fig.~\ref{fig:torque_density_integrated_torque} we focus on the orbit-averaged torque characteristics, examining their variation as a function of planetary eccentricity $e$. In panels (a) and (b) we plot the radial torque density $dT/dr$ and the integrated torque $T(r)$ respectively. The different coloured lines denote the progression of $e$ from low (cool colours) to high (hot colours) values. Meanwhile in panel (c), we have decomposed the separate torque contributions in accordance with the method outlined in Section \ref{subsection:circular_theory}. The black, red and blue points denote $T_\textrm{net}$, $T_\textrm{L}$ and $T_\textrm{C}$ respectively as a function of $e$. 

We immediately see the strong change in the radial structure of the averaged torque profile in panel (a). For low $e$, $dT/dr$ matches onto the circular profile with an approximately anti-symmetric structure in the inner and outer disc. As $e$ increases, the torque density begins to invert. By $e=0.05$, $dT/dr$ has levelled out as a stationary point of inflection in the vicinity of the guiding radius. As the eccentricity reaches the transonic value of $e = 0.06$ and beyond, the torque density slope becomes negative at $r=a$ so that positive torque is now injected into the inner disc and negative torque in the outer disc. Accordingly, the extrema of $dT/dr$ either side of the guiding radius have reversed in sign and drifted slightly away from $r = a$. One also notices the enhanced asymmetry in the profile for larger $e$, which leads to a more pronounced torque differential between the inner and outer disc. For the supersonic eccentricities, before the torque density levels out far from $r=a$, we also see that $dT/dr$ overshoots zero, which can be interpreted as a radially shifted and reduced version of the original peak from the low-$e$ profiles.

These features are clearly borne out in the integrated torque profile $T(r)$ shown in panel (b). Recall that this quantity is the one-sided torque integrated from $r=a$, as defined according to equation \eqref{eq:one_sided_torque}, and so the net torque is easily extracted as the difference between the plateaued profile in the inner and outer disc. For low eccentricities, the saturated $T(r)$ values (measured far from $r=a$) have only a slight asymmetry leading to a small net torque on the disc and planet. As $\tilde{e}$ crosses unity, we see a rapid evolution in these plateaus as they both shift downwards before settling down into a new regime as $\Tilde{e}\rightarrow 2$. In this supersonic regime, the asymmetries are more pronounced and $T(r_\textrm{out}) < T(r_\textrm{in})$ so the torque differential has actually reversed in sign, as was first suggested by \cite{PapaloizouLarwood_2000}.

In the panel (c) we have decomposed the full torque $T_\textrm{net}$ into the Lindblad and corotation contributions. We see that as $e$ exceeds 0.06, $T_\textrm{net}$ begins to turnover and shortly afterwards becomes negative. Towards the end of our computational range, the net torque appears to be saturating. The Lindblad torque closely traces this behaviour, meanwhile the corotation contribution weakly decays\footnote{This weakening of the corotation torque may be reminiscent of the shrinking of the horseshoe librating region around high-eccentricity planets in particle discs \citep{Namouni1999,Rafikov2003}.} with eccentricity while maintaining the same sign. As a result, for large eccentricities the net torque is dominated by the Lindblad component.

\subsection{Non-coorbital corotation resonant features}
\label{subsection:non-coorbital resonances}

In our torque decomposition pipeline we consider the corotation contributions from every $(m,l)$ combination. Generally $m \neq l$ and hence the pattern speed $\omega_{ml}\neq \Omega_\textrm{p}$. Therefore, in addition to the usual coorbital resonances associated with circular planetary orbits, there are also a host of non-coorbital corotation resonances at $r_{ml}$ such that $\omega_{ml}=\Omega(r_{ml})$. These are typically much weaker and subdominant compared with the coorbital and Lindblad torques. Nonetheless they can still be seen in our radial angular momentum profiles. In principle, since any $l/m$ ratio can excite a corotation response in a different location, the disc is densely populated with lots of little jumps which can superimpose (for commensurable modal ratios) or be finely spaced to give a complicated substructure.

\begin{figure}
    \centering
    \includegraphics[width=\columnwidth]{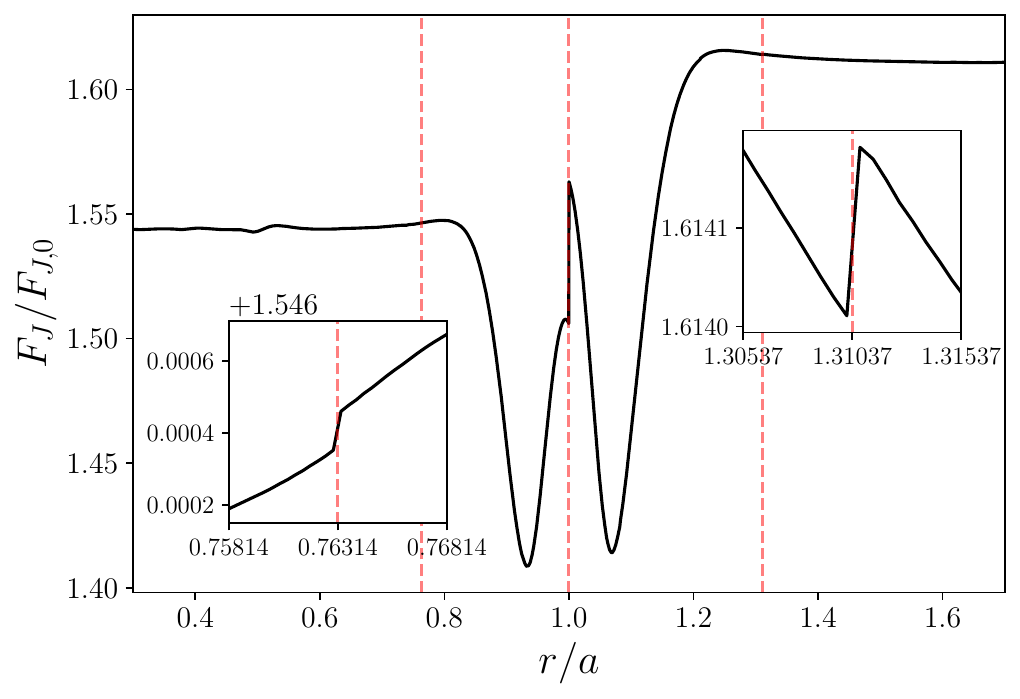}
    \caption{Radial profile of the angular momentum flux for a $q=p=0.0$, $h_\textrm{p}=0.06$ disc and a planet with transonic eccentricity of $e=0.06$. The red dashed lines denote, moving from left to right, the $l=3$  $m=2$, coorbital, and $l=2$ $m=3$ corotation resonances. The non-coorbital resonances are examined more closely via the zoomed in panels, revealing the small amplitude jumps of $F_J$ indicative of the corotation torques at these locations. See Section \ref{subsection:non-coorbital resonances} for more details.}
    \label{fig:non_coorbital_features}
\end{figure}

For illustration, in Fig.~\ref{fig:non_coorbital_features} we plot $F_{J}$ for the case of $q=p=0.0$, $h_\textrm{p}=0.06$ and a transonic eccentricity of $e=0.06$. The dashed red lines at $0.76314 a$ and $1.31037a$ denote the corotation resonances for the $(m,l) = (2,3)$ and $(3,2)$ modes respectively, whilst the middle dashed line emphasizes the usual coorbital location. The inset panels zoom in to the profiles at the non-coorbital locations. The expected coorbital angular momentum flux jump is clearly visible and dominates the linear corotation torque with an amplitude of $\sim 0.05 F_{J,0}$. However, the inset panels show clear discontinuities also at non-coorbital corotation resonances, with much smaller amplitudes of order $10^{-4} F_{J,0}$. Recalling that the Fourier potential expansion goes as $e^{|l-m|}$, non-coorbital modes are typically driven by the weaker perturbing components. This is compounded by the fact that the non-coorbital pattern speeds shift the resonance further away from the planet where the potential is weaker. These combined effects severely limit the influence of the non-coorbital contributions to the net torque. Nevertheless, collectively they are still very important to account for and ignoring them introduces significant errors. Indeed, for the case shown in Fig.~\ref{fig:non_coorbital_features} the coorbital $m=l$ corotation torques contribute 0.052 $F_{J,0}$ to the net corotation torque. Meanwhile, all the other non-coorbital $m \neq l$ terms contribute 0.009 $F_{J,0}$. This is $\sim 17\%$ of the total corotation torque and should not be disregarded.

The continuous radial profile of the AMF in the vicinity of the coorbital location is also notably different for this eccentric case, compared with the circular examples shown in Fig.~\ref{fig:circular_morphology_torques}. In the circular cases, the AMF increases as one moves radially away from $r=a$. However, for this eccentric example the AMF actually decreases towards minima either side of the coorbital radius, before increasing again and plateauing further away from the planet. For globally isothermal discs recall that $dF_J/dr = dT/dr$ (see Section \ref{subsection:circular_wake_amf}) so the reversal in the flux profile is consistent with the torque behaviour seen in panels (a) and (b) of Fig.~\ref{fig:torque_density_integrated_torque} for $e = 0.06$.

This decrease in the AMF and integrated torque about $r=a$ can be attributed to the fact that the eccentric orbit introduces pattern speeds which no longer move with the planetary mean motion. This results in some inner Lindblad resonances (which contribute a negative torque density) moving exterior to $a$ and outer Lindblad resonances (which contribute a positive torque density) moving towards the interior disc. Physically, \cite{PapaloizouLarwood_2000} interpret this effect as the fact that the planet trails the local gas flow at apocenter and leads it at pericenter, so the excited wake in these regions tends to pull on the planet and exert a torque in the opposite sense of the usually repulsive one-sided Linblad torques. The radial AMF structure here justifies this interpretation since the minimum turning points of $F_J$, either side of $r=a$, occur in the vicinity of the planetary apo/pericenters.

\subsection{Torque trends with $q$ and $p$}
\label{subsection:torque_trends}

\begin{figure*}
    \centering
    \includegraphics[width=0.9\textwidth]{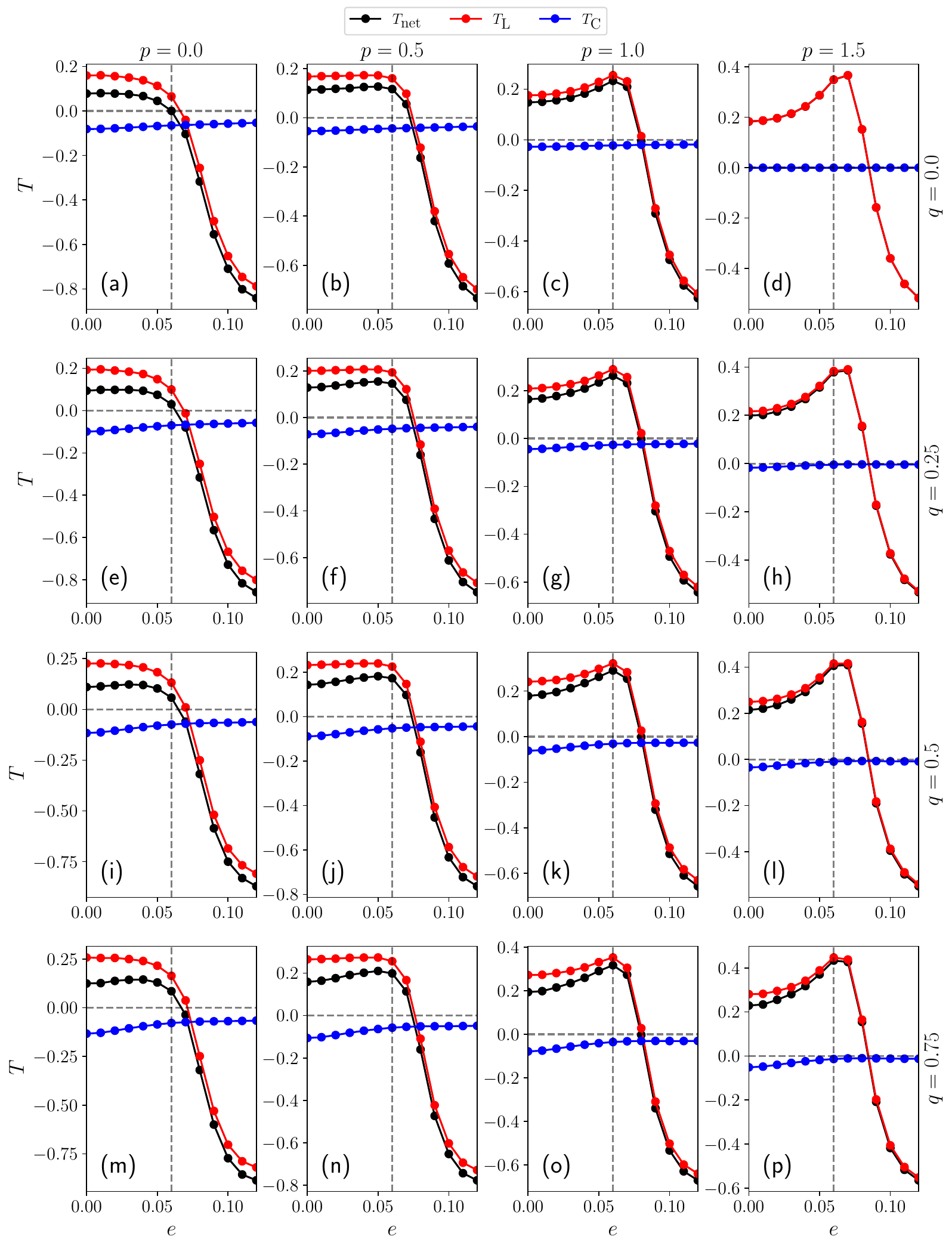}
    \caption{The torque decomposition as a function of eccentricity across a grid of disc properties $(q,p)$ for $h_\textrm{p}=0.06$. Each panel corresponds to a different $(q,p)$ with columns being labelled by $p\in[0.0,0.5,1.0,1.5]$ and rows being labelled by $q\in[0.0,0.25,0.5,0.75]$. The black, red and blue data points denote the value of $T_\textrm{net}$, $T_\textrm{L}$ and $T_\textrm{C}$, respectively, as a function of $e$, where the torques are all normalised by the characteristic value $F_{J,0}$. The vertical dashed line shows the transonic value of $e$ and the horizontal dashed line divides the torque reversal about $T$=0.}
    \label{fig:torque_eccentricity_h_0.06}
\end{figure*}

In Fig.~\ref{fig:torque_eccentricity_h_0.06} we now plot the variation in the different torque components as a function of eccentricity for each of the runs across our $p$ and $q$ grid. Here we will first focus on the case of $h_\textrm{p} = 0.06$ which best allows us to probe the transonic transition. These figures have the same structure as the rightmost panel of Fig.~\ref{fig:torque_density_integrated_torque}. From this grid we can extract some qualitative trends. 

In all cases the broad behaviour is similar to that discussed for the fiducial run discussed in Section \ref{subsection:radial_torque_structure} -- namely the torque reversal occurring at $\tilde{e}\sim 1$. This robustly extends the prediction of \cite{PapaloizouLarwood_2000} to a broader range of disc profiles. Furthermore, the Lindblad torque always tracks the behaviour of the net torque, with slight adjustments due to $T_\textrm{C}$ which presents a more gradual variation with eccentricity. In particular, for the globally isothermal disc in panel (d), where $q=0.0$ and surface density exponent $p=1.5$, the corotation contribution becomes 0 for all eccentricities as the vortensity gradient vanishes (see Section \ref{subsection:circular_wake_amf}). Contrastingly, as one considers higher $q$ exponents in the case of a uniform surface density disc, the effect of corotation is maximised for panel (m), with the blue points tracing a slightly curved profile with largest absolute magnitude for low $e$ before levelling off at larger $e$. It should be noted that even for discs where the vortensity gradient vanishes, the linear corotation torque is still non-zero in panels (h), (l) and (p) since the non-uniform temperature structure introduces an entropy gradient corotation contribution which is often over-looked \citep[e.g.][]{PaardekooperEtAl_2010}.

A more pronounced qualitative feature is observed in the net and Lindblad torques as we push towards steeper surface density profiles with $p=1.5$ ----- the emergence of a bump around the transonic $e\approx h_\mathrm{p}$ value. This bump presents an enhancement in the positive torque enacted on the disc before the torque turns-over and reverses for larger $e$. For low eccentricities we see the same qualitative trends with $q$ and $p$ as for the circular planet case (see Fig.~\ref{fig:circular_torques}). That is, the net torque enacted on the disc increases with both $q$ and $p$. Meanwhile, for the high eccentricity regime where the net torque is negative, we see that as $p$ increases, $T_\mathrm{net}$ still increases and therefore reduces the absolute magnitude of the reversed planet-disc torque. Furthermore, as $q$ increases $T_\textrm{net}$ actually decreases very slightly for fixed $p$, although this subtle trend is difficult to discern in the figure.

So far we have focused on the particular case of $h_\textrm{p} = 0.06$. However, our wider parameter space exploration also covers the torque behaviour for $h_\textrm{p} = 0.08$ and $h_\textrm{p} = 0.1$. In Appendix \ref{app:parameter_space}, we present figures analogous to Fig.~\ref{fig:torque_eccentricity_h_0.06} which display the torque components as a function of $e$ across $(q,p)$ space for these additional values of $h_\textrm{p}$. Since $h_\textrm{p}$ is larger in both of these cases, the same eccentricity range extends to smaller values in $\tilde{e}$ and hence the planet remains subsonic for larger values of $e$ than before. In these plots the transonic critical line has shifted to the right and the torque profiles appear qualitatively similar to before but stretched horizontally, so that the torque reversals and bumps associated with $\Tilde{e}\sim 1$ now occur for larger $e$. Furthermore, we notice that as $h_\textrm{p}$ is increased, the magnitude of the torque for any particular $e$ also increases, as might be anticipated from the behaviour for circular planets shown in Fig.~\ref{fig:circular_torques}.

\subsection{Orbital evolution timescales}
\label{subsection:orbital_evolution}

For practical applications, we are interested in converting the torques discussed earlier into the orbital evolutionary timescales, which we do now. In particular, the change in angular momentum of the planet needs to be translated into changes in its semi-major axis and eccentricity. Indeed, the angular momentum of the planet is given by
\begin{equation}
\label{eq:AM_planet}
    L = M_\textrm{p}\sqrt{GM_* a (1-e^2)} = M_\textrm{p}\Omega_\textrm{p} a^2 \sqrt{1-e^2} ,
\end{equation}
such that 
\begin{equation}
\label{eq:L_a_e_evolution}
    \frac{1}{L}\frac{dL}{dt} = \frac{1}{2a}\frac{da}{dt}-\frac{e}{1-e^2}\frac{de}{dt}.
\end{equation}
Following \cite{IdaEtAl_2020} we define the inverse orbital evolution timescales
\begin{equation}
\label{eq:inverse_timescales}
    \tau_{L}^{-1} = -\frac{1}{L}\frac{dL}{dt} , \quad \tau_{a}^{-1} = -\frac{1}{a}\frac{da}{dt} , \quad \tau_{e}^{-1} = -\frac{1}{e}\frac{de}{dt},
\end{equation}
such that equation \eqref{eq:L_a_e_evolution} becomes
\begin{equation}
\label{eq:timescale_relation}
    \tau_{L}^{-1} = \frac{1}{2}\tau_{a}^{-1}-\frac{e^2}{1-e^2}\tau_{e}^{-1}.
\end{equation}
Note, that in previous literature the notation for the angular momentum damping rate is often given by $\tau_\textrm{m}^{-1}$ instead \cite[e.g.][]{PapaloizouLarwood_2000,IdaEtAl_2020}, where the subscript `\textrm{m}' misleadingly implies migration rate. We intentionally change this convention as the migration behaviour is more correctly described by the semi-major axis evolution rather than the angular momentum for the eccentric planetary orbit. These rates can be related to the torque quantities which we have previously computed. Recalling that $T_\textrm{net}$ is the net torque enacted on the disc, the back-reaction onto the planet gives $dL/dt = -T_\textrm{net}$ so
\begin{equation}
\tau_{L}^{-1} = \frac{T_\textrm{net}}{L}.
\end{equation}
Of course the planetary torques are equal to the magnitude of the disc torques but opposite in sign, in order to conserve the net angular momentum of the system. For the semi-major axis evolution, first consider the impact of individual modes separately. Each mode excites a response in the disc with a fixed pattern speed $\omega_{ml}$. This in turn corresponds to a gravitational potential perturbation acting back on the planet. A uniformly rotating perturbing potential is known to conserve the Jacobi integral $E_\textrm{J} = E-\omega_{ml}L$ where $E = -GM_* M_\textrm{p}/(2a)$. Therefore taking the temporal derivative leads to an energy evolution $dE/dt = -\omega_{ml}T_{\textrm{net},ml}$ which can be related to the semi-major axis evolution as
\begin{equation}
    \tau_{a}^{-1} = \frac{2}{M_\textrm{p }a^2 \Omega_\textrm{p}^2}\sum_{m,l} \omega_{ml}T_{\textrm{net},ml},
\end{equation}
upon summing all modal contributions. Finally, we obtain the eccentricity damping timescale $\tau_{e}^{-1}$, by simply rearranging equation \eqref{eq:timescale_relation}.

\begin{figure*}
    \centering
    \includegraphics[width=0.9\textwidth]{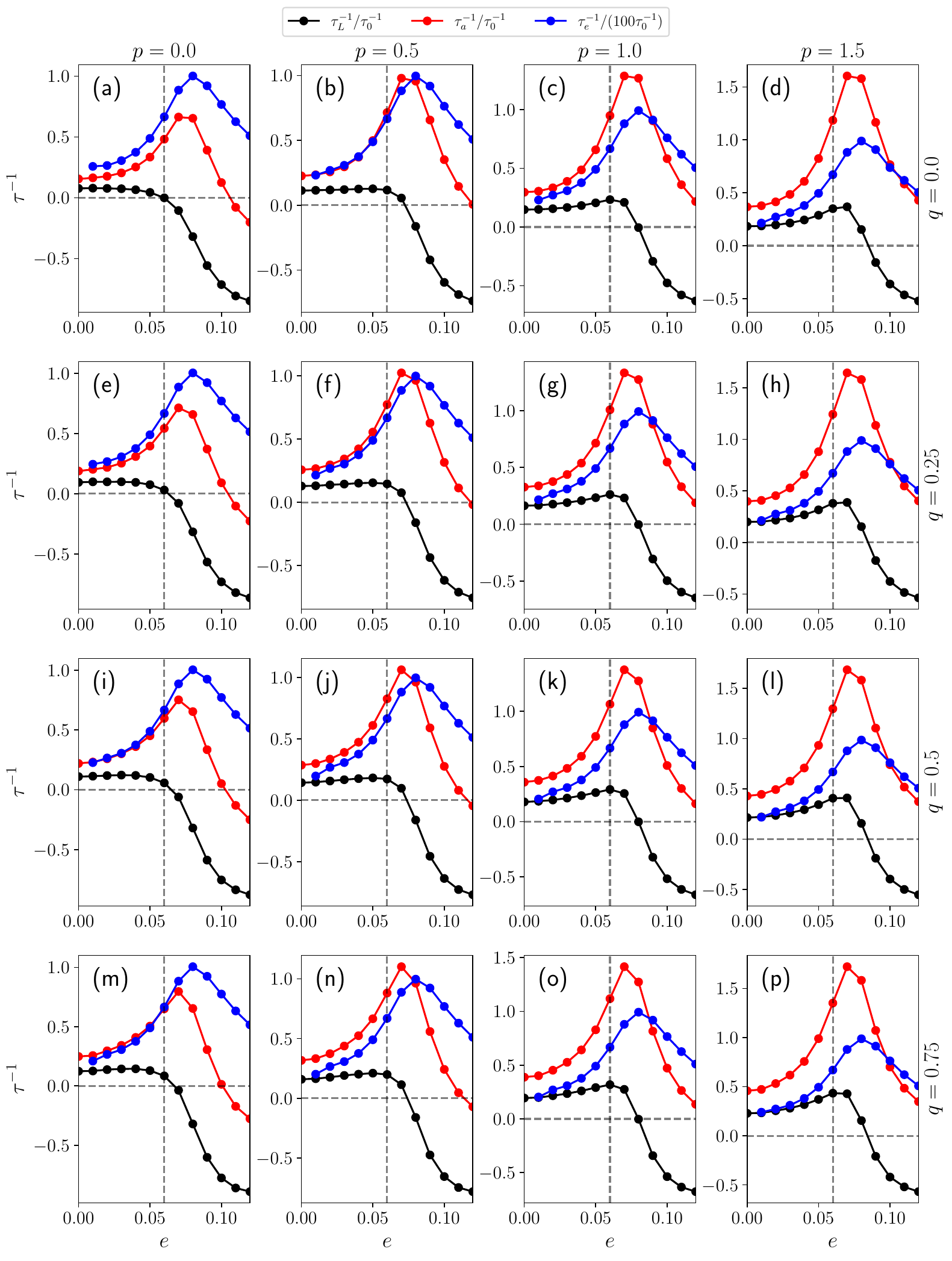}
    \caption{The orbital evolution timescales as a function of eccentricity across a grid of disc properties $(q,p)$ for $h_\textrm{p}=0.06$, as per Fig.~\ref{fig:torque_eccentricity_h_0.06}. The black, red and blue data points denote the value of $\tau_{L}^{-1}/\tau_{0}^{-1}$, $\tau_a^{-1}/\tau_{0}^{-1}$ and $\tau_\textrm{e}^{-1}/(100\tau_{0}^{-1})$, respectively, as a function of $e$. The vertical dashed line shows the transonic value of $e$ and the horizontal dashed line divides the torque reversal about $T$=0.}
    \label{fig:eccentric_timescales_h_0.06}
\end{figure*}

Following this procedure we are able to extract the characteristic evolution timescales and plot them as functions of eccentricity in Fig.~\ref{fig:eccentric_timescales_h_0.06}. Akin to Fig.~\ref{fig:torque_eccentricity_h_0.06}, the grid of panels show the results for each $(q,p)$ combination of disc properties, with fixed $h_\textrm{p} = 0.06$. The black, red and blue curves now correspond to $\tau_{L}^{-1}$, $\tau_{a}^{-1}$ and $\tau_{e}^{-1}$ respectively. Note that the angular momentum and semi-major axis inverse timescales are normalised by the characteristic inverse timescale
\begin{equation}
\label{eq:tau_0_inv}
    \tau_{0}^{-1} = \frac{F_{J,0}}{M_\textrm{p}\Omega_\textrm{p} a^2}.
\end{equation}
Meanwhile the eccentricity damping occurs much faster and is instead scaled by 100$\tau_{0}^{-1}$, allowing us to plot the results on the same axes.

The profiles reveal some interesting behaviour. Since $\tau_{L}^{-1}$ is directly proportional to the net torque, these tracks simply follow the same trends as $T_\textrm{net}$ discussed in Section \ref{subsection:torque_trends}; namely the torque reversal is once again obvious for supersonic values of $e$. However, despite planetary angular momentum being increased in this regime, this increase notably does not directly translate into an immediate reversal of the direction of planetary migration, as emphasized recently by \cite{IdaEtAl_2020}. Indeed, since the angular momentum budget is dependent on both $a$ and $e$, a sufficiently fast damping of eccentricity can allow the angular momentum to increase even if $a$ is decreasing. As a result, in all panels, for eccentricities where the black line first reverses in sign, the semi-major axis evolution timescale remains positive and therefore still corresponds to inwards migration. In fact, inspection of the red points in Fig.~\ref{fig:eccentric_timescales_h_0.06} show that $\tau_{a}^{-1}$ actually peaks just beyond $\tilde{e} = 1$, underlining that the damping of semi-major axis is most efficient in this transonic regime. We see that this peak is strongest in panel (p) where the disc has the steepest surface density and temperature profiles. These peaks in $\tau_{L}^{-1}$ also roughly coincide with peaks in $\tau_{e}^{-1}$ which suggest a much more rapid eccentricity damping occurring on the order of 100 times faster (recall our choice of scalings). Beyond these peaks, towards higher eccentricity, both $\tau_{a}^{-1}$ and $\tau_{e}^{-1}$ turnover and start to decrease. 

For discs with shallow surface density gradients, we observe that $a$ migration can in fact eventually reverse direction within the range of $e$ probed by our study (e.g. see the high eccentricity results in panel (m)). Such high-$e$ orbital expansion was first proposed by \cite{PapaloizouLarwood_2000} for the specific case study of a disc with $p=1.5$ but not found in subsequent studies (see discussion in Section \ref{subsection:previous_results}). Indeed we find that outwards migration actually occurs most readily in uniform surface density discs and that the methodology of \cite{PapaloizouLarwood_2000} produces order unity discrepancies which render it unreliable for predicting when such reversals actually take place. For the larger values of $p$, extrapolation of our results suggests that outwards migration may also be possible for even larger values of $e$ although the curves do appear to be levelling out. Meanwhile, $\tau_{e}^{-1}$ also turns over as the eccentricity damping becomes less efficient for higher $e$. Although our pipeline does not probe very large eccentricities, the downwards trends in $\tau_{e}^{-1}$ perhaps hint at eventual eccentricity pumping for highly supersonic values of $e$, as suggested by the gas dynamical friction approaches of \cite{BuehlerEtAl_2024} and \cite{ONeill2024}. However, it should be noted that these previous studies adopt a homogeneous and static gaseous background, which does not self-consistently incorporate the background Keplerian shear flow. In all parameter space panels, the shape of the eccentricity damping rate appears similar, suggesting that it is relatively insensitive to the choice of $q$ and $p$, in keeping with the findings of previous results \citep{Artymowicz_1993,TanakaEtAl_2002}.

Once again, we also present the extended parameter space results for $h_\textrm{p} = 0.08$ and $0.1$ in Appendix \ref{app:parameter_space} where the shift in transonic eccentricity stretches the profiles towards higher $e$ in an approximately self-similar fashion. This means that the enhancements in eccentricity damping and inwards migration rate shift to higher values of eccentricity. Indeed, for the case of $h_\textrm{p} = 0.1$, $e$ is not extended to large enough values to capture the overturn in these peak rates. As per the increase in torques with larger $h_\textrm{p}$ seen in Fig.~\ref{fig:torque_eccentricity_h_0.06}, here we also note that the orbital evolution rates increase with $h_\textrm{p}$.

%% file: Sections/5_discussion.tex
\section{Discussion}
\label{sec:discussion}

Our methodology provides a robust and self-consistent exploration of the 2D linear torque across various disc parameters and eccentricities. In the previous sections we presented the torque and evolutionary timescale results at some discrete points in the space of disc parameters. In the following Sections \ref{subsection:fitting_functions} and \ref{subsection:migration_tracks} we build upon these results, while providing some additional comparisons and discussion in Sections \ref{subsection:previous_results} and \ref{subsection:caveats}.

\subsection{Fitting functions}
\label{subsection:fitting_functions}

Although the shape of the torque and migration timescale profiles presented in Section \ref{sec:eccentric_torques} appears relatively simple, our attempts to derive simple fitting functions across the three-dimensional $(q,p,h_\mathrm{p})$ parameter space has proven challenging. Indeed, the shape of the $T_\textrm{net}$ curve in Fig.~\ref{fig:torque_eccentricity_h_0.06} is suggestive of a hyperbolic tangent superimposed with a (skewed) Gaussian component. Whilst this ansatz can well describe the curve for a specific panel, there is not a simple relationship linking the fitting parameters across the whole grid as a function of $q$ and $p$ (never-mind the variation associated with the wider data set as a function of $h_\textrm{p}$). This emphasizes the complexity of the general problem and also cautions against the use of simple fitting formulae. 

Instead we have opted to our make data available to community and provide our full data cube of torques and evolutionary timescales in the supplementary material which can be easily read in and interpolated by users for their own purposes. The details of this data are described more in Appendix \ref{app:parameter_space}.

\subsection{Migration tracks}
\label{subsection:migration_tracks}

To illustrate the usage of our results, we computed some exemplar tracks for the orbital evolution of eccentric planets as a function of time. When doing this, it needs to be remembered that we have computed the torques assuming that the planetary position is held fixed and hence so too are the local disc properties, which in particular scale the dimensional torque through the characteristic flux quantity $F_{J,0}$. However, as the planet moves in $a$, the local aspect ratio will change in general even for a fixed disc profile with constant $(q,p)$ and $h_\mathrm{p}$ corresponding to the initial planetary semi-major axis. To account for this variation of $h$ as the planet migrates, we linearly interpolate our data cube of orbital evolution rates onto the instantaneous values for $(a,e,h(a))$ and the local disc properties. This procedure is detailed more in Appendix \ref{app:parameter_space} where we describe how to access and manipulate our full data cube, with exemplar reading and plotting routines available in the supplementary material. 

\begin{figure}
    \centering
    \includegraphics[width=\columnwidth]{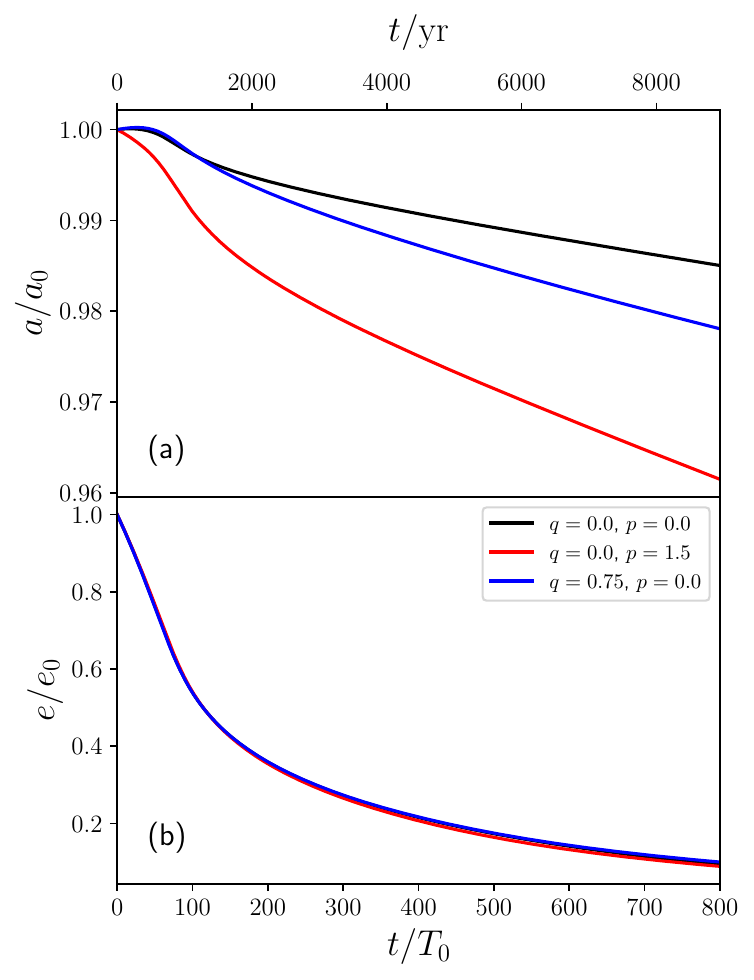}
    \caption{Migration tracks for an Earth mass planet embedded in a disc at initially $a_0 = 5$AU, where $\Sigma(5\textrm{AU}) = 355$ g cm$^{-2}$ and $e_0 = 0.12$. \textit{Upper panel}: semi-major axis evolution. \textit{Lower panel}: eccentricity evolution. Time is measured in orbital periods $T_0$ at $a_0$. The black, red and blue curves denote disc profiles which adopt $(q,p) = (0.0,0.0)$, $(0.0,1.5)$ and $(0.75,0.0)$ respectively.}
    \label{fig:1d_migration}
\end{figure}

With these details in mind, we consider the orbital evolution of an  $M_\mathrm{p}=M_\oplus$ protoplanet around a $M_\odot$ central star. The protoplanet starts at a semi-major axis of $a_0 = 5$ AU with an initial eccentricity $e_0 = 0.12$ (the highest probed in our work) and is embedded in a disc with local surface density $\Sigma(5\textrm{AU}) = 355$ g cm$^{-2}$ and aspect ratio $h(a_0) = 0.062$. Note that we choose this starting aspect ratio to be slightly larger than the minimum value of $h_\textrm{p}=0.06$ so that as the planet migrates inwards, its environment remains within the range of parameters surveyed in Section \ref{sec:eccentric_torques}. This planetary mass is suitably sub-thermal for our adopted aspect ratio, where $M_\mathrm{p}/M_\textrm{th} = 0.0126$. The resulting orbital evolution tracks are plotted in Fig.~\ref{fig:1d_migration} as a function of time, where we see the evolution of $a/a_0$ in the upper panel and $e/e_0$ in the lower panel over 800 orbital timescales $T_0$ (measured with respect to the initial planetary position) or equivalently $9\times 10^3$ yr. Tracks computed for the three different combinations of $(q,p) = (0.0,0.0)$, $(0.0,1.5)$ and $(0.75,0.0)$ are plotted as the black, red and blue lines respectively.
%

We see that over the explored time interval, the semi-major axis changes very little whilst the eccentricity varies quite dramatically. Orbital elements vary the most at the very beginning of the tracks, as the planet progresses through a range of eccentricities. In the uniform density discs (black and blue curves), we observe a brief spell where the planet migrates outwards in accordance with the semi-major axis drift rate reversals seen in panels (a) and (m) of Fig.~\ref{fig:torque_eccentricity_h_0.06}. However, this trend quickly reverses as the eccentricities decrease and the typical inwards migration resumes -- in fact peaking around the transonic value of $e$. The three curves split into distinct paths as the varying disc properties separate the migration rates, with the steepest surface density discs leading to the quickest inwards drift. Meanwhile, the eccentricity paths almost overlap suggesting that the eccentricity damping is only weakly sensitive to the details of the disc profile, as suggested by the quantitatively similar $\tau_e^{-1}$ curves seen across all panels in Fig.~\ref{fig:eccentric_timescales_h_0.06}. 

\begin{figure}
    \centering
    \includegraphics[width=\columnwidth]{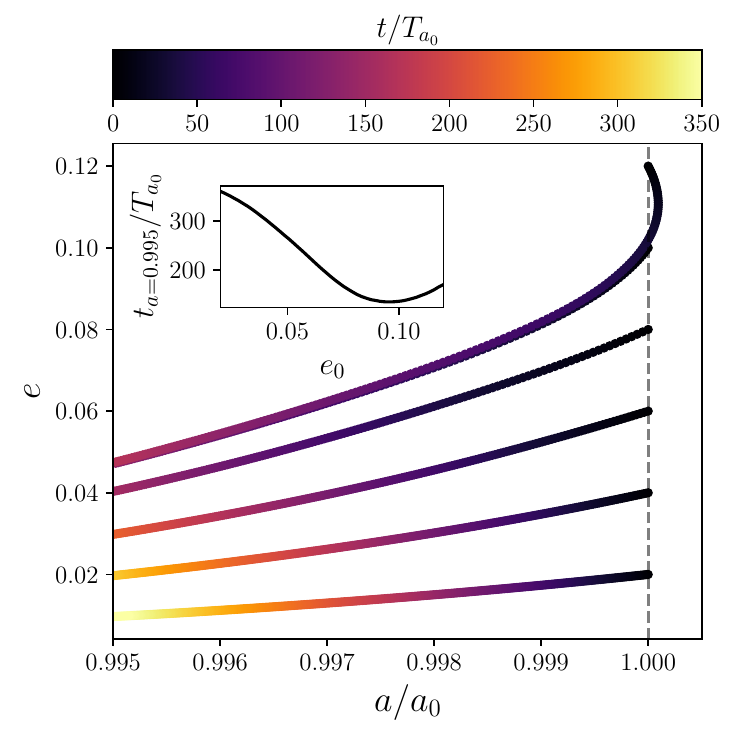}
    \caption{Migration tracks in $(a,e)$ phase space for planets in a uniform surface density, globally isothermal disc with $a_0 = 5$AU and $\Sigma(5\mathrm{AU}) = 355$ g cm$^{-2}$. The different tracks correspond to different initial eccentricities $e_0$ and the colour of each curve indicates the time in units of the orbital period at $a_0$. The dashed vertical line denotes the initial semi-major axis of the planet. The inset shows the time it takes for each track to reach $a=0.995a_0$ as a function of $e_0$.}
    \label{fig:a_e_tracks}
\end{figure}

Another way to visualise this evolution is to consider tracks in $(a,e)$ space as shown in Fig.~\ref{fig:a_e_tracks}. Here, we once again assume the same system properties as for the black curve in Fig.~\ref{fig:1d_migration} wherein the disc has a uniform temperature and surface density profile. However, now the $M_\mathrm{p}=M_\oplus$ planet is initialised at $5$ AU (along the vertical dashed grey line) with different eccentricities ranging from $e_0=0.02$ to $0.12$ in steps of $0.02$. The migration is followed for each of these initial conditions for 350 orbital timescales at the starting semi-major axis ($\sim 3.9\times 10^3$ yr). The time along each track is visualised by the colour. Notably, for the most eccentric initial condition we see a brief period of outwards migration with continued eccentricity damping. This trend rapidly reverses once the planetary eccentricity has diminished, and the planet then starts migrating inwards, closely overlying the curve for the $e=0.10$. 

The initial eccentricity clearly segregates the time at which the different curves reach a fixed value of $a$. For example, the inset panel shows the time at which tracks of varying $e_0$ cross $a/a_0 = 0.995$. We see that this time reaches a minimum for starting eccentricities $e_0$ are around the peak of $\tau_a^{-1}$, as seen in Fig.~\ref{fig:eccentric_timescales_h_0.06}. Meanwhile, it takes longer to reach $0.995a_0$ for the tracks with lower and higher $e_0$, for which the migration in less efficient (or even reverses sign).

\subsection{Comparison with previous prescriptions}
\label{subsection:previous_results}

We now provide a detailed comparison of our eccentric planet-disc coupling results with past studies. As pointed out by \cite{IdaEtAl_2020}, the existing literature is complicated by a host of studies which probe different disc regimes and employ various methodologies and approximations. Often these approaches are stitched together in a heuristic fashion to motivate torque fitting formulae over a wider parameter space. Inevitably such a procedure has led to a variety of possible planet migration characteristics depending on which model is adopted. For these reasons, here we will focus on the comparison with two previous studies which are most relatable to our current approach -- namely they investigate 2D discs and consider torque behaviour across a range of eccentricities, in particular pushing towards supersonic values. 

\begin{figure}
    \centering
    \includegraphics[width=\columnwidth]{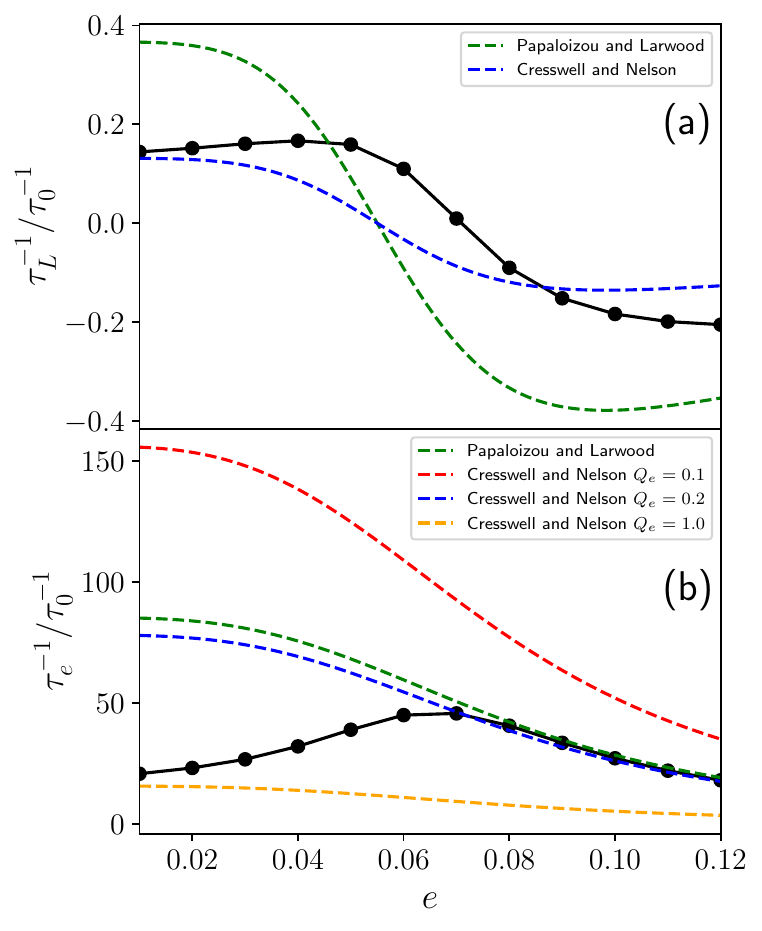}
    \caption{Migration rates for a disc with $q = 1.0$, $p = 1.5$, $h_\textrm{p} = 0.05$ and $b=0.5$. Our data is plotted as the black points and connecting points. \textit{Upper panel (a)}: Comparison of angular momentum damping rates $\tau_{L}^{-1}$. Green dashed curve is the PL00 prediction given by the equation \eqref{eq:tau_m_inv_PL00}. Blue dashed curve is the CN06 prescription given by the equation \eqref{eq:tau_m_inv_cn}. \textit{Lower panel (b)}: Comparison of eccentricity damping rates $\tau_e^{-1}$. Green dashed curve is for PL00 prescription, see equation \eqref{eq:tau_e_PL00}. The red, blue and orange dashed lines plot the CN06 prescription given by the equation \eqref{eq:tau_e_inv_cn} for $Q_e =$ 0.1, 0.2 and 1.0 respectively.}
    \label{fig:tau_m_comparison}
\end{figure}

Firstly, we summarise the approach of \cite{PapaloizouLarwood_2000} (hereafter PL00) who considered discs with $q=1.0$, $p=1.5$ and aspect ratio $h_\textrm{p} = 0.05$ (uniform across the disc for the adopted $q$). Note, that this choice of surface density profile is often targeted by previous studies in order to suppress the complicating effects of corotation resonances (see Section \ref{subsubsec:corotation}). However, although this choice of $p$ removes the leading order corotation term associated with vortensity gradients, it does not remove the entropy gradient contributions mentioned in Section \ref{subsection:torque_trends}. Whilst they also extract the full $(m,l)$ Fourier decomposition of the perturbing potential, instead of solving for the wave mode structure numerically they exploit the modified WKB formula of \cite{Artymowicz_1993}, as presented earlier in equation \eqref{eq:L_WKB}. They provide fits for the resulting angular momentum and eccentricity damping rates to be
\begin{equation}
    \label{eq:tau_m_inv_PL00}
    \tau_{L,\textrm{PL00}}^{-1} = 7.33 \left(\frac{b}{0.5}\right)^{-1.75}h_\textrm{p}\frac{1-(\Tilde{e}/1.1)^4}{1+(\Tilde{e}/1.3)^5}\tau_0^{-1} ,
\end{equation} 
\begin{equation}
    \label{eq:tau_e_PL00}
    \tau_{e,\textrm{PL00}}^{-1} = 4.26 \left(\frac{b}{0.5}\right)^{-2.5}\left(1+\frac{1}{4}\tilde{e}^3\right)^{-1}h_{\textrm{p}}^{-1}\tau_0^{-1},
\end{equation}
as a function of softening  parameter $b$, aspect ratio $h_\mathrm{p}$ and, crucially, the scaled eccentricity $\tilde{e}$. This early work clearly predicts a torque reversal when $\tilde{e}\sim 1.1$, in keeping with our findings also. Furthermore, the eccentricity damping rate decreases $\propto \tilde{e}^3$ for high eccentricities. The PL00 orbital evolution rates (\ref{eq:tau_m_inv_PL00})-(\ref{eq:tau_e_PL00}) are illustrated in Fig.~\ref{fig:tau_m_comparison} with green dashed curves.

Secondly, \cite{CresswellNelson_2006} (hereafter CN06) performed 2D, hydrodynamic simulations for a range of planetary eccentricities also embedded in discs with $q=1.0$, $p=1.5$ and $h_\textrm{p} = 0.05$. They directly measured the angular momentum and eccentricity damping rates which were compared to the predictions of PL00. They state that their measured angular momentum damping rates at low $e$ are a factor of $\sim 3$ slower than the PL00 prediction, whilst the high $e$ rates are a factor of $\sim 3/2$ slower. Meanwhile, they find good agreement with the predicted eccentricity damping timescale for high $e\gtrsim 0.08$ but report a clear discrepancy for $e\lesssim 0.05$ where PL00 again give a rate roughly 3 times faster. CN06 also emphasize the sensitivity to the softening parameter and find that a value of $b\sim 0.5-0.6$ gives good agreement with the 3D analytical results of \cite{TanakaEtAl_2002} for the migration rate. 


Informed by these simulations, they combine in a somewhat ad-hoc fashion 3D analytical calculations presented by \cite{TanakaEtAl_2002} and \citet{TanakaWard_2004} with the eccentricity dependencies found by PL00. Their suggested angular momentum damping rate prescription is 
\begin{equation}
    \label{eq:tau_m_inv_cn}
    \tau_{L,\textrm{CN06}}^{-1} = \frac{2.7+1.1 p}{2} h_\textrm{p} \frac{1-(\Tilde{e}/1.1)^4}{1+(\Tilde{e}/1.3)^5}\tau_0^{-1} ,
\end{equation}
which apparently gives very good agreement with their 2D simulation measurements for low $e$ and decent agreement at large $e$. Meanwhile, their final eccentricity prescription is more questionable and given by
\begin{equation}
\label{eq:tau_e_inv_cn}
    \tau_{e,\textrm{CN06}}^{-1} = 
    \frac{0.78}{Q_e} \left(1+\frac{1}{4}\tilde{e}^3\right)^{-1} h_{\textrm{p}}^{-1}\tau_0^{-1} ,
\end{equation}
where $Q_{e}\sim 0.1$ was chosen by \cite{CresswellNelson_2006} to approximately model their simulation results. The CN06 orbital evolution rates (\ref{eq:tau_m_inv_cn})-(\ref{eq:tau_e_inv_cn}) are also shown in Fig.~\ref{fig:tau_m_comparison}, for different $Q_e$ in panel (b).

Given that CN06 visually demonstrates good agreement between the simulations and PL00 for high $e$, we find that the choice of $Q_e \sim 0.2$ brings equations \eqref{eq:tau_e_PL00} and \eqref{eq:tau_e_inv_cn} into better accordance, see blue dashed curve in Fig.~\ref{fig:tau_m_comparison} (b). Such a prescription still disagrees significantly with the low $e$ simulation results for $\tau_{e}^{-1}$, as per the comparison with the PL00 prescription discussed previously. Therefore CN06 implement a constant rate below $e<0.08$ to limit the damping rate. In fact, if one instead adopts $Q_e = 1$ (orange dashed curve in Fig.~\ref{fig:tau_m_comparison}(b)) then the low eccentricity limit of equation \eqref{eq:tau_e_inv_cn} becomes  equivalent to the analytical result proposed by \cite{TanakaWard_2004} for small eccentricity damping rates in a 3D disc. It might be hoped that a low-$e$, 2D setup with $b \sim 0.5$ would then agree better with this analytical result, akin to the aforementioned match found for $\tau_{L}^{-1}$ between the softened 2D simulations of CN06 and the 3D migration rates presented by \cite{TanakaEtAl_2002}.

To compare our methodology with these previous studies, we use our pipeline to extract the torque for a comparable disc with $q=1.0$, $p=1.5$, $h_\textrm{p} = 0.05$ and $b = 0.5$. We then compute $\tau_{L}^{-1}$ and $\tau_e^{-1}$ as a function of planetary eccentricity and compare them with the literature prescriptions in Fig.~\ref{fig:tau_m_comparison}. We plot our results as the black data points connected by a solid line. In panel (a), we compare with the PL00 prediction \eqref{eq:tau_m_inv_PL00} and the CN06 prescription \eqref{eq:tau_m_inv_cn}. The CN06 result seems to qualitatively agree with our results, especially at low $e$, although some discrepancies are obvious in the transonic and high-$e$ regimes. By extension, our results should also qualitatively agree well with the 2D simulations of CN06. On the other hand, the PL00 curve deviates from our calculation by a factor of $\sim 2.5$ in the low-$e$ limit and $\sim 1.7$ in the high-$e$ limit. This discrepancy is reminiscent of similar differences found between PL00 and the simulations of CN06. Hence our method aligns better with the numerics and justifies the use of our global linear theory. Namely, it emphasizes the importance of calculating the full global structure of the excited wave modes if one cares about order unity differences, rather than simply the localised WKB approach.

In the lower panel (b), we instead compare the eccentricity damping rates. Our results are once again plotted as black dots connected by the solid line, which clearly exhibits a characteristic enhancement around the transonic value of $e$. This enhancement in the eccentricity damping timescales (as well as semi-major axis damping in Fig.~\ref{fig:eccentric_timescales_h_0.06}), is robustly found in simulations of eccentric planets embedded in discs \citep[e.g. CN06,][]{Cresswell2007}. However, this effect is typically not captured by orbital migration fitting functions presented elsewhere in the literature \citep{IdaEtAl_2020}. For example, the PL00 prescription (equation \eqref{eq:tau_e_PL00}) is plotted as the dashed green line. In keeping with the numerical comparison of CN06, this gives very good agreement for supersonic values $e$. However, for low values of $e$ the PL00 result gives a damping rate, which is faster than ours by a factor of $\sim 4$. This echoes the discrepancy found by CN06 who reported the eccentricity damping in their simulations to be a factor of $\sim 3$ slower than PL00 and thus in much better alignment with our results. The CN06 model given by equation \eqref{eq:tau_e_inv_cn} is also over-plotted for three different values of $Q_{e} = (0.1,0.2,1.0)$ as the red, blue and orange lines respectively. The $Q_e = 0.2$ result give close correspondence with the PL00 result and shows the expected good agreement with our results at high $e$. Meanwhile, the $Q_e = 1.0$ result gives better agreement for the lower eccentricity end. Our result, which is in closer agreement with the simulations, seems to be an interpolation of these two limiting cases (i.e. may be modelled using $e$-dependent $Q_e$, dropping as $e$ increases). It should be emphasized that whilst these previous prescriptions perform well in limited eccentricity intervals, only our self-consistent result smoothly connects across the transonic regime and captures both qualitative and quantitative consistency with past simulations.

Finally, it should be noted that our method also yields a semi-major axis damping rate which remains positive for the adopted $p$ and all examined eccentricities and therefore does not exhibit outwards migration. Although not plotted here, we find that $\tau_a^{-1}$ is levelling off towards a positive value beyond $e>0.12$, which is consistent with the fact that CN06 report no outwards migration of the planet even for large values of eccentricity. This is in stark contrast with the results of PL00 which predicted expansion of the planetary orbit for transonic eccentricities (as seen in Fig. 1 of \cite{IdaEtAl_2020}).

\subsection{Caveats of our approach}
\label{subsection:caveats}

Our approach provides a self-consistent calculation of the linear torque across a wide disc parameter space and range of eccentricities. However, we acknowledge that this scope comes at the expense of a number of compromises which we now mention.

Firstly, we should remember that we are calculating the response for a 2D disc. The process of vertical averaging motivates a softened potential, which introduces the somewhat arbitrary $b$ parameter. This is typically tuned by comparison of 2D results with 3D simulations \citep[e.g.][]{CresswellNelson_2006}. Notably, \cite{Cresswell2007} and \cite{DAngeloLubow_2010} performed both 2D and 3D simulations showing that the 2D torque can be brought into qualitative agreement with the 3D value for specific choices of $b$.  However, the variation of torque with disc characteristics $p$ and $q$ is also sensitive to the exact value of $b$ chosen. More recent numerical simulations of \cite{JimenezMasset_2017a} corroborate these results in the locally isothermal case.

There has also been analytical work extending the planet-disc interaction to three dimensions. In particular, \cite{TanakaEtAl_2002} solved for the 3D wave modes and calculated the resulting migration rate for a circular planet, avoiding the need for a softening parameter. Recently this has been extended to the locally isothermal case \citep{TanakaOkada_2024} which compares favourably with the simulations discussed above. However, although they considered an eccentric perturber with small $e$ in \cite{TanakaWard_2004}, this did not allow them to probe the supersonic regime of key interest in this work.

Another notorious complication is the treatment of the corotation torque. We have limited our attention to the case of the linear corotation torque. However, the evanescent waveform launched in the vicinity of corotation resonances \citep{GoldreichTremaine_1979} means that the angular momentum is deposited locally and might modify the background disc structure. This develops as a nonlinear \textit{horseshoe} drag \citep[e.g.][]{Ward_1991, PaardekooperEtAl_2010} wherein fluid elements exchange angular momentum with the planet as they undergo librating orbits in the vicinity of corotation. If there is sufficient viscosity in the disc to communicate away the deposited angular momentum flux, it is reasoned that the linear corotation result can still be a good approximation. Indeed, \cite{OgilvieLubow_2003} showed that the classic linear result can be recovered for non-coorbital eccentric corotation resonances when the perturbing potential strength relative to the dissipation is small. We can therefore speculate that as the eccentricity of the perturber increases and the the non-coorbital resonances become more important, the linear corotation torques may dominate over the nonlinear horseshoe contributions. 

Whilst these caveats will change some of the quantitative details for 3D discs which permit corotation nonlinearities, the qualitative behaviour we find seems to be borne out in previous simulations. We encourage comparison with targeted future simulations of 2D and 3D locally isothermal eccentric planet-disc interactions to further test our results.





%% file: Sections/6_conclusions.tex
\section{Conclusions}
\label{sec:conclusion}

In this paper we have conducted a systematic parameter sweep of 2D, locally isothermal eccentric planet-disc interactions. We have directly solved the global linearised wave equations for a variety of disc temperature profiles, surface densities and aspect ratios using the methodology described in \cite{FairbairnRafikov_2022}. Our direct numerical calculation of modes circumvents the approximations associated with the classical WKB calculations. The resulting wave structure allows us to compute the net torque, which we further decompose into the Lindblad and linear corotation contributions. We perform this for planetary eccentricities ranging from 0.0 to 0.12. In particular, this allows us to probe planetary orbital evolution as the planet begins to move supersonically relative to the gas. The ratio $\tilde{e} = e/h_\textrm{p}$ is the critical measure of this transition and hence our run with the lowest value of $h_\textrm{p} = 0.06$ allows us to interrogate this transition best. 

As $\tilde{e}$ exceeds unity, the torque density undergoes clear changes in its radial profile, with the inner and outer disc contributions reversing in sign. This is notably accompanied by a net torque reversal where angular momentum is now injected into the planetary orbit. However, a net torque increase does not necessarily yield outwards migration: within the range of parameter space investigated, the outwards migration only happens for the highest eccentricities in thin (low $h_\mathrm{p}$) discs with shallow surface density gradients (small $p$). More generally, in the transonic regime (at $\tilde e\sim 1$) we see a strong enhancement in eccentricity and semi-major axis damping rates. The rapid eccentricity damping in the transonic case (roughly $10^2$ times faster than the semi-major axis damping) should efficiently circularize high-$e$ planets and eventually force them to migrate inwards. 

We have also bench-marked our approach by performing an extensive sweep through the space of relevant disc parameters, the results of which were compared with previous analytical and numerical investigations of eccentric planet-disc interaction. Our results appear to be consistent with the existing numerical calculations, although future 2D and 3D simulations across a range of disc models would be desirable to test this further. This will also help address the caveats intrinsic to our 2D, linear methodology which includes the uncertain softening parameter as well as the neglect of nonlinear horseshoe corotation contributions.

Finally, we make our full results freely available in the supplementary materials for future testing and for implementation by the community in e.g. the exoplanetary population synthesis studies.

%% file: Sections/acknowledgements.tex
\section*{Acknowledgements}

CWF would like to acknowledge insightful conversations with Josh Brown and Gordon Ogilvie. This research was supported by the W. M. Keck Foundation Fund, the Adler Family Fund, and the Sivian Fund (CWF), as well as the STFC grant ST/T00049X/1 and the IAS (RRR). 

%% file: Sections/app_convergence.tex
\section{Torque convergence}
\label{app:torque_converegence}

Extending our torque calculation to eccentric discs requires some care to ensure suitably converged results. Whilst in \paper1 we performed a convergence study which used the spatial wake morphology as the convergence metric, here we are interested in a robust quantitative convergence of the planet-induced torque. As discussed in Section \ref{subsubsection:convergence} we must consider convergence of $T_\textrm{net}$ in both $m$ and $l$. In particular the extension to higher $e$ runs requires the inclusion of a higher degree of `off-diagonal' modes with larger values of $|l-m|$. 

\begin{figure}
    \centering
    \includegraphics[width=\columnwidth]{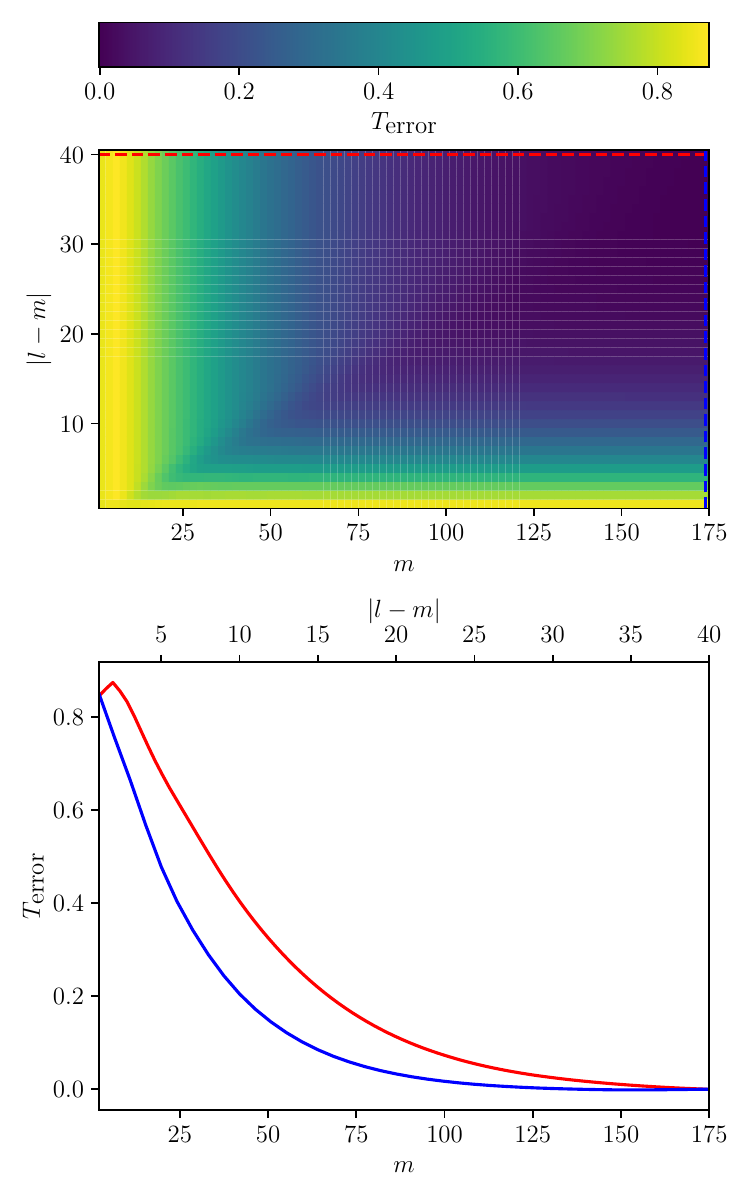}
    \caption{\textit{Upper panel}: heat map plotting the torque error, $T_\textrm{error}$, as a function of the number of included modes up to some upper limits in $m$ and $|l-m|$. $T_\textrm{error}$ is measured with respect to the limiting value in the upper right corner where $m=175$ and $|l-m|=40$. \textit{Lower panel}: plots slices through the heat map showing the convergence in $m$ for fixed $|l-m|=40$ (\textit{red line}) and $|l-m|$ for fixed $m=175$ (\textit{blue line}).}
    \label{fig:convergence}
\end{figure}

Fig.~\ref{fig:convergence} characterises this convergence behaviour for the case of a supersonic eccentricity orbit $e=0.12$ ($\tilde e=2$), in a disc with $q=p=0.0$, $h_\textrm{p}=0.06$ and $b=0.3$. In the upper panel we plot a heat map which shows the error in the net torque $T_\textrm{error}$, where the x-axis denotes the number of included $m$ modes whilst the y-axis denotes the maximum order taken in $|l-m|$. Then $T_\textrm{error}$ is computed with respect to a reference value for which the maximum number of modes is included -- i.e. in this case the upper right of the plot where $m_\textrm{max}=175$ and $|l-m|_\textrm{max}=40$. One sees that there is a diagonal line tracing the steepest descent in error. By eye, this immediately suggests that $m > 100$ and $|l-m|>20$ is desired for decent results here. 

We also examine slices through this convergence space. In the lower panel the red line shows the convergence in $m$ whilst holding $|l-m|=40$ (corresponding to the red, dashed line in the upper panel). Meanwhile the blue line shows the convergence in $|l-m|$ whilst $m=175$ is fixed (corresponding to the blue dashed line in the upper panel). Both exhibit a steep, exponential drop-off before beginning to level out. The clear flattening of the blue curve confirms that an upper value of $|l-m| = 40$ is more than enough. Meanwhile the red curve follows a shallower descent which has nearly levelled out but still retains a shallow gradient at the upper value of $m=175$. This slope is small however and suggests that additional values of $m$ will only make small differences. 

Such a gradient descent is suggestive of some automated procedure for mode finding, where the $m$ and $|l-m|$ limits of the modes included in the calculation keep getting extended until some threshold in the slope is reached. Care must be taken for certain cases where the convergence structure might not be as simple as shown in this heat map and local minima can emerge. Such an automated procedure is left to future streamlined pipelines. Instead, inspection of a variety of runs (in particular those probing the most numerically challenging eccentricities) gives us confidence that $m_\textrm{max}=175$ and $|l-m|_\textrm{max}=40$ is sufficient for our purposes. In fact, for subsonic values of $e$, this pipeline demonstrates that a more truncated expansion in $m$ is sufficient and we employ $m_\textrm{max}=150$ for such $e$. Pushing to highly supersonic values of $e$, with $\Tilde{e}>2$, causes a much slower convergence which becomes numerically limiting for our present study.

%% file: Sections/app_spurious_modes.tex
\section{Puzzling modal solution}
\label{app:m1q1}

\begin{figure}
	\centering
	\includegraphics[width=\columnwidth]{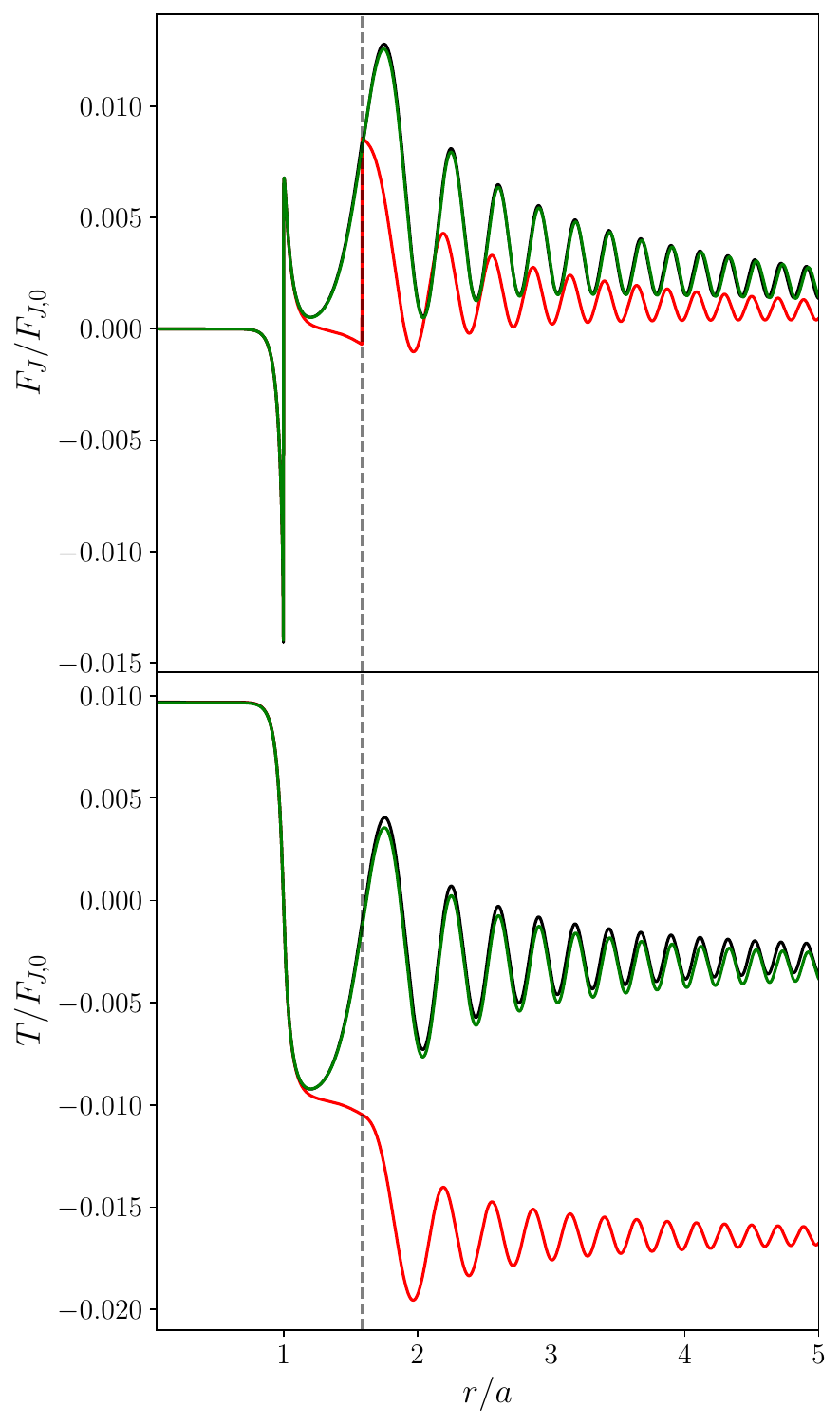}
	\caption{The $m=1$ response for a disc with $h_\textrm{p} = 0.06$, $b = 0.3$ and $p = 0.0$ for a circular planet with $e = 0.0$. \textit{Upper panel}: angular momentum flux. \textit{Lower panel}: integrated torque. The vertical dashed line shows the position of the outer Lindblad resonance. The red lines correspond to $q=1.0$ whilst the black and green lines (almost overlapping) correspond to $q=1.005$ and $q=0.995$ respectively.}
 \label{fig:circular_q1_torque}
\end{figure}

During the testing of our pipeline for circular orbits with $e = 0.0$, we also extended our runs to larger values of $q$, which highlighted one surprising result. Precisely at the value $q=1.0$ and $p=0.0$ we found that the solution is not well behaved for the $m=1$ mode. In Fig.~\ref{fig:circular_q1_torque} we plot the angular momentum profile for this mode in the upper panel and the integrated radial torque in the lower panel. The red line denotes the special value of $q=1.0$ whilst the black and green lines denote $q=1.005$ and $q=0.995$ respectively. In all cases a uniform surface density is chosen such that $p=0.0$. The dashed grey vertical line denotes the outer Lindblad resonance where there is clearly a discontinuous jump in $F_{J}$ for the $q=1.0$ case, whilst the lines remains smooth for $q = $ 0.995 and 1.005. This spurious behaviour is also echoed in the integrated torque profile which branches off to lower values beyond the outer Lindblad resonance, unlike for values of $q=1.005$ and $q=0.995$ which curve back upwards. By checking a variety of $q$ values approaching 1.0 from above and below we see that this transition in behaviour is sudden and therefore nonphysical. This is seconded by efforts to integrate the equations from a different initial radius which also gave very different looking curves in the case of $q=1.0$, suggesting there is some numerically ill-posed point across the outer Lindblad resonance here. 

Subsequently, when extending our analysis to the full parameter space of eccentric runs, we also noticed some slight deviations in the smooth trends seen in the torque and orbital migration timescale data. This was also traced back to a small handful of spurious $m=1$ modes which present a behaviour similar to that discussed above. Again, this misbehaviour is highly localised in parameter space and slight shifts in the values of $q$ and $p$ characterizing disc properties remove the problem. Throughout this paper we tackle these rare anomalies by actively removing the troublesome modes or infinitesimally shifting the parameter space point to a nearby value. So far we have been unable to track down the origin of this behaviour, which will be deferred for future work.

%% file: Sections/app_parameter_space.tex
\section{Extended Parameter Space}
\label{app:parameter_space}

\begin{figure*}
    \centering
    \includegraphics[width=0.9\textwidth]{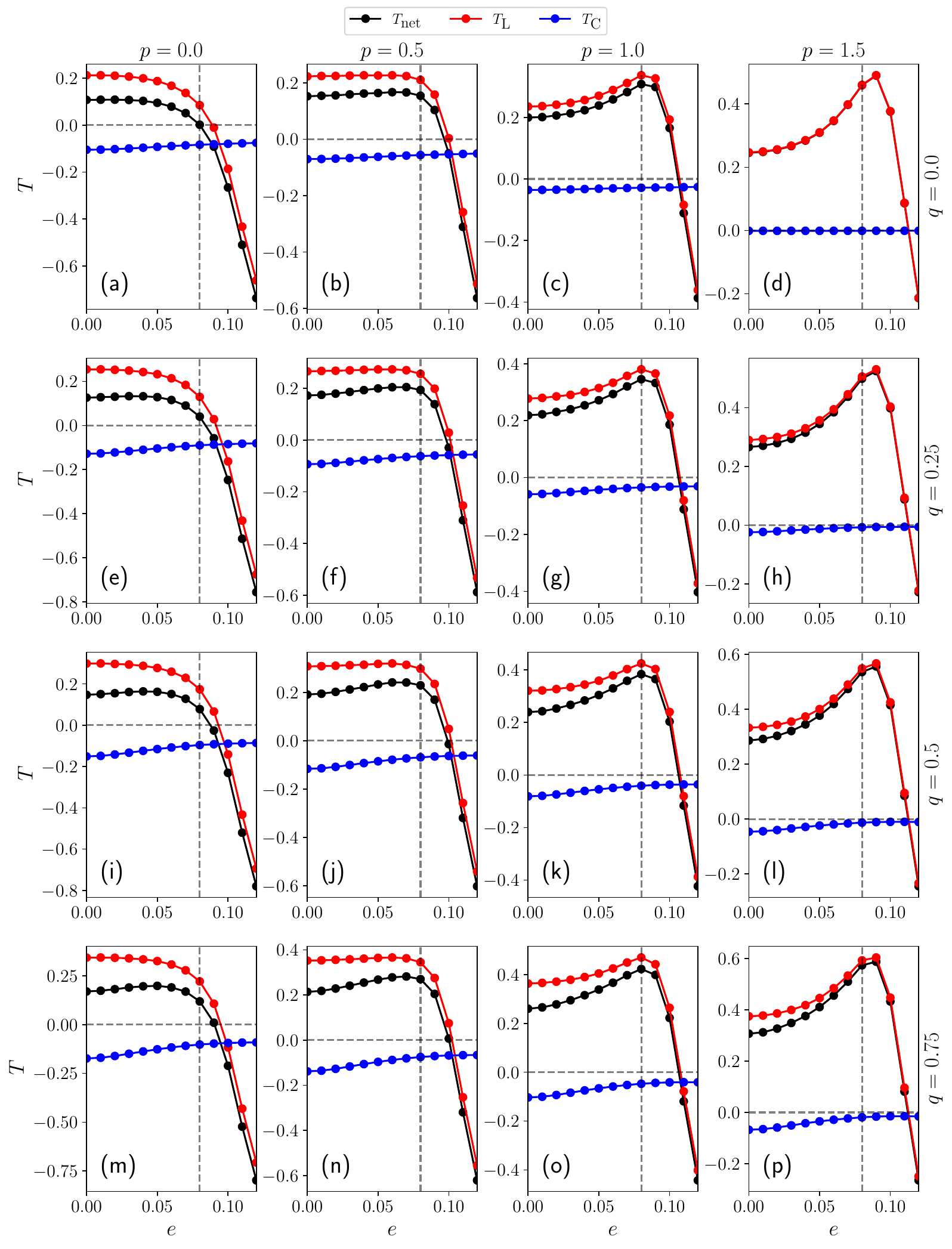}
    \caption{As per Fig.~\ref{fig:torque_eccentricity_h_0.06} but for $h_\textrm{p}=0.08$.}
    \label{fig:torque_eccentricity_h_0.08}
\end{figure*}

\begin{figure*}
    \centering
    \includegraphics[width=0.9\textwidth]{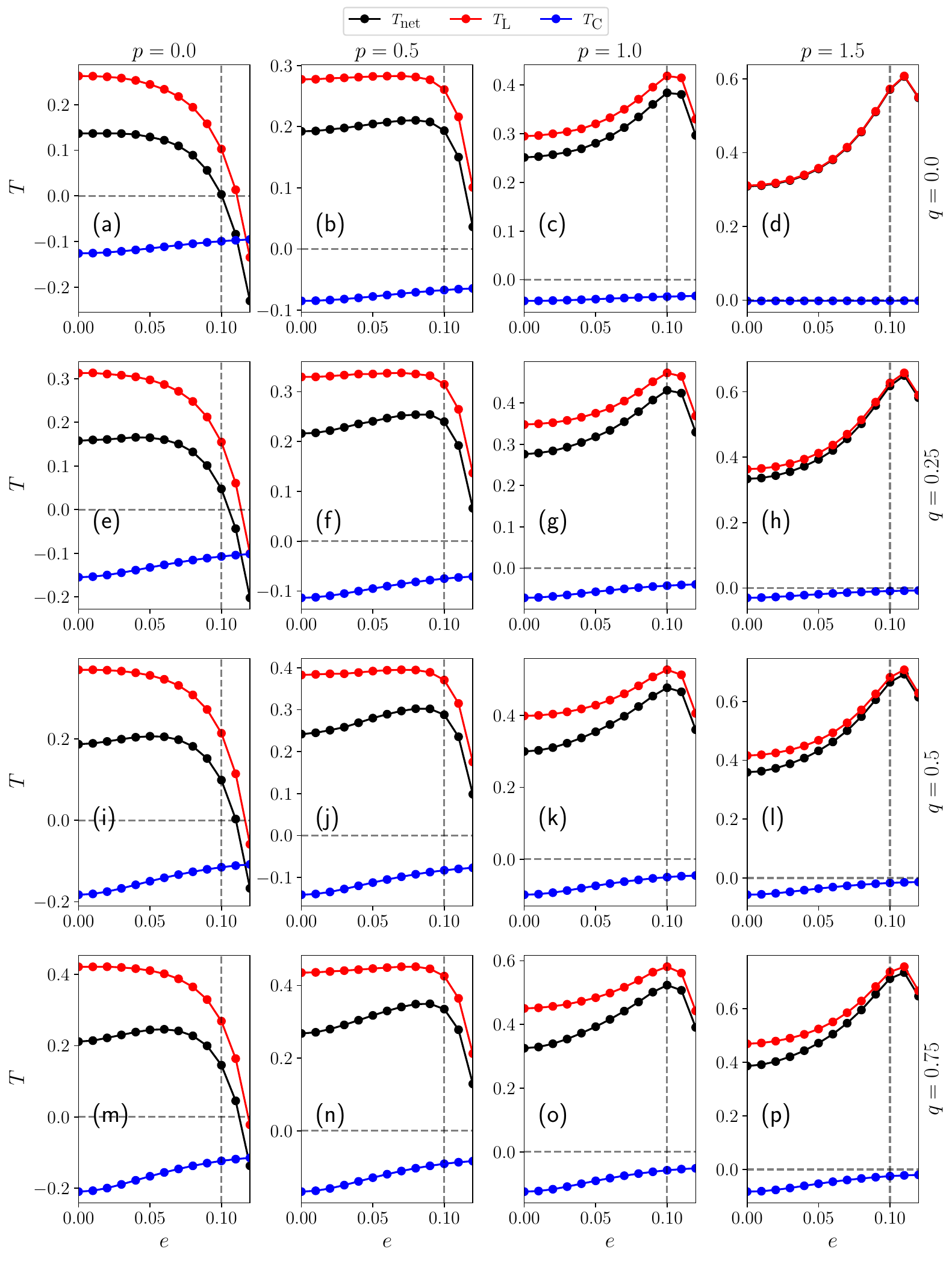}
    \caption{As per Fig.~\ref{fig:torque_eccentricity_h_0.06} but for $h_\textrm{p}=0.10$.}
    \label{fig:torque_eccentricity_h_0.1}
\end{figure*}

\begin{figure*}
    \centering
    \includegraphics[width=0.9\textwidth]{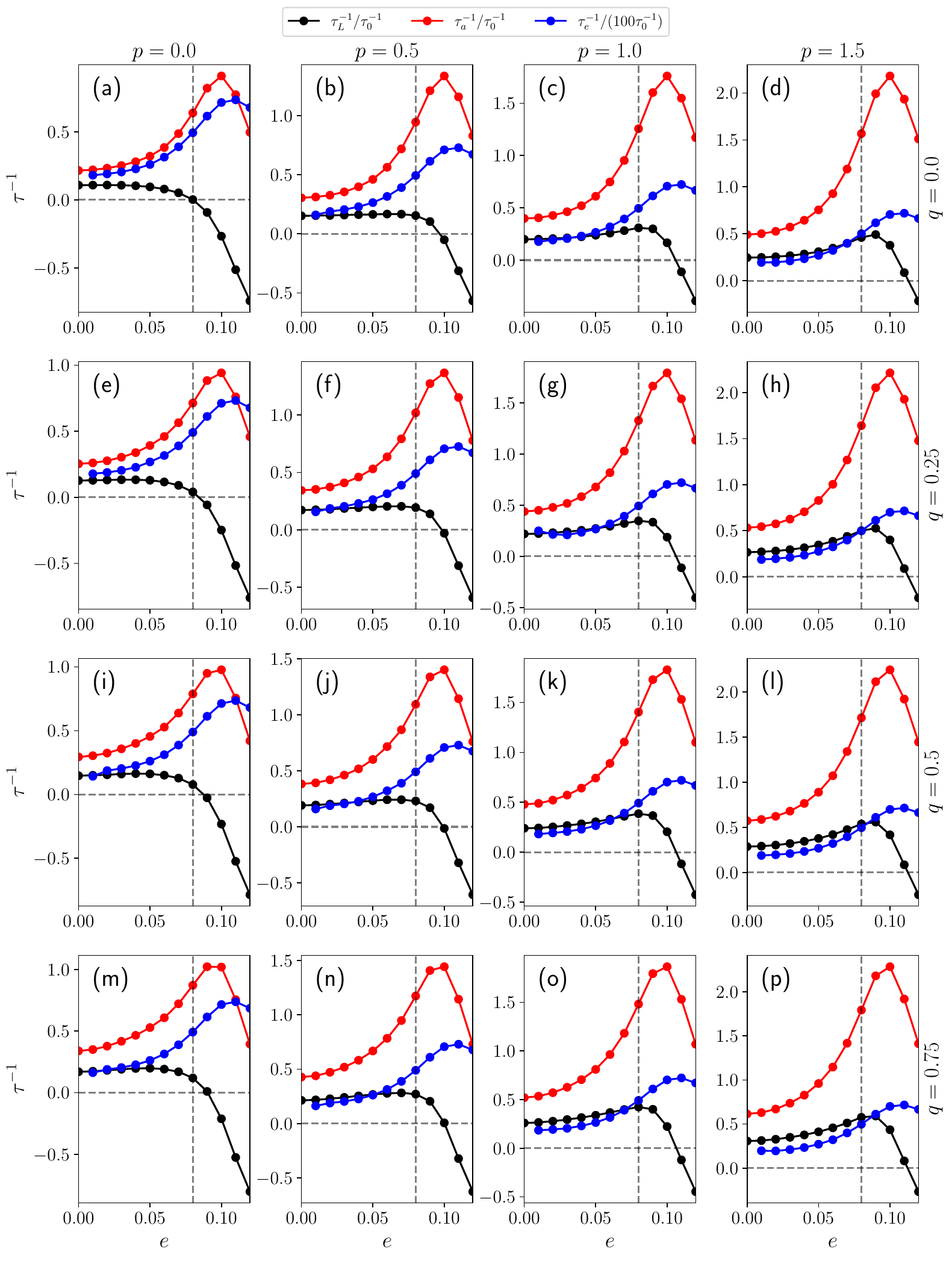}
    \caption{As per Fig.~\ref{fig:eccentric_timescales_h_0.06} but for $h_\textrm{p}=0.08$.}
    \label{fig:eccentric_timescales_h_0.08}
\end{figure*}

\begin{figure*}
    \centering
    \includegraphics[width=0.9\textwidth]{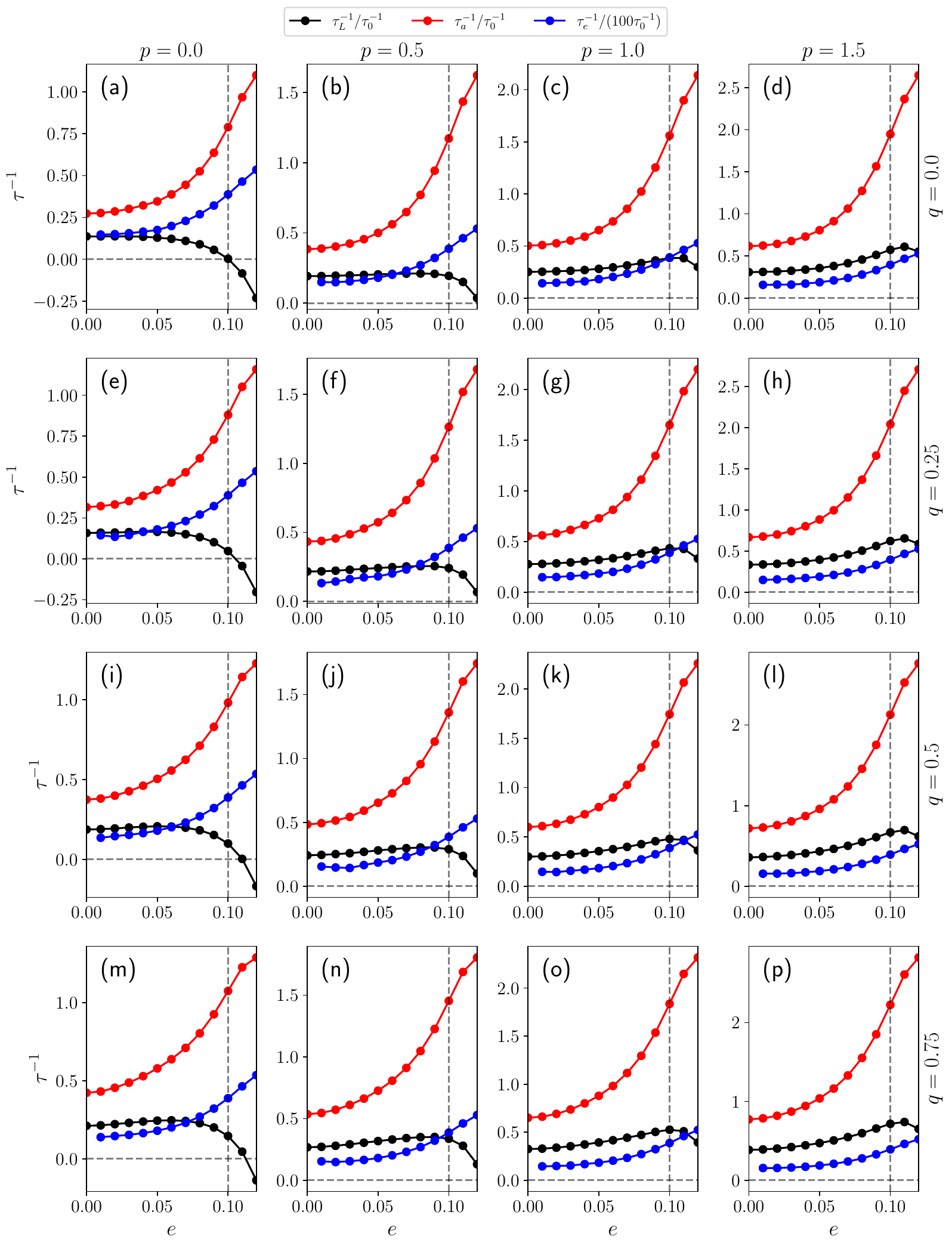}
    \caption{As per Fig.~\ref{fig:eccentric_timescales_h_0.06} but for $h_\textrm{p}=0.10$}
    \label{fig:eccentric_timescales_h_0.1}
\end{figure*}

In Section \ref{sec:eccentric_torques} we focused our discussion on the  $h_\textrm{p} = 0.06$ discs, while varying other disc parameters. This is the lowest value of $h_\textrm{p}$ in our main parameter sweep and therefore allows us to probe the largest values of the ratio $\tilde{e} = e/h_\textrm{p}$. We also computed the torque and migration timescale results for higher values of $h_\textrm{p} = 0.08$ and $0.1$, allowing us to explore the variation of torque with aspect ratio. In Figs.~\ref{fig:torque_eccentricity_h_0.08} and \ref{fig:torque_eccentricity_h_0.1} we plot the different torques as a function of eccentricity across $(q,p)$ space, akin to Fig.~\ref{fig:torque_eccentricity_h_0.06}. Then in Figs.~\ref{fig:eccentric_timescales_h_0.08} and \ref{fig:eccentric_timescales_h_0.1} we do the same for the migration rates, as per the fiducial Fig.~\ref{fig:eccentric_timescales_h_0.06} in the main text. 

Furthermore, we make our full data cube available to the community as supplementary material to the journal and also as a public repository online (\href{https://github.com/callumf8/Eccentric-planet-disc-interactions.git}{https://github.com/callumf8/Eccentric-planet-disc-interactions.git}). Accessing and manipulating this data cube is described more in the accompanying \texttt{README.md} file. Furthermore, we have provided an exemplar jupyter notebook script \texttt{read\_data.ipynb}, which performs the reading in of the $(q,p,h_\mathrm{p})$ data cube, interpolation onto intermediate regions of parameter space and demonstrates how to calculate migration tracks as shown in Fig.~\ref{fig:1d_migration}. Whilst some extrapolation outwith the bounds of our study is perhaps justified, one should be careful not to extend it too far as our results demonstrate the complicated and sensitive dependence on the underlying disc parameters and planetary eccentricities.
